\documentclass[aps, pra, preprint, groupedaddress, amsfonts,
               amsmath, amssymb, showpacs, nofootinbib]{revtex4-1}
\usepackage{microtype}
\usepackage{graphicx}
\usepackage{epstopdf}
\usepackage[utf8]{inputenc}
\usepackage[T1]{fontenc}
\usepackage[usenames,dvipsnames]{xcolor}
\usepackage{hyperref}

\usepackage{amssymb}

\def\idt{{i} \partial _t}

\def\bra#1{\hbox{$\langle #1 |$}}
\def\ket#1{\hbox{$| #1 \rangle $}}
\def\scp#1#2{\hbox{$\langle #1 | #2 \rangle $}}

\def\part#1#2{{\partial #1 \over \partial #2}}

\def\Rtoinfty{\vtop{ \tiny \hbox to 1cm{\hss $\longrightarrow$ \hss} \hbox to 1cm{\hss $ R\to \infty $\hss}}}
\def\TwoVbox#1#2#3#4{\hbox to #1\hsize{\hss \hsize=#1\hsize
  \vbox to #2\vsize{\hsize=0.48\hsize \vfill #3 \vfill} \hss
  \vbox to #2\vsize{\hsize=0.48\hsize \vfill #4 \vfill} \hss}}
\def\d{\mathop{\raise0pt\hbox{\rm d}}}
\def\i{\mathop{\raise0pt\hbox{\rm i}}}

\begin{document}
\title{Calculation of Energy Loss in Antiproton Collisions with Many-Electron Systems using Ehrenfest's Theorem}

\author{Hans J\"urgen L\"udde}
\email[]{luedde@itp.uni-frankfurt.de}
\affiliation{Center for Scientific Computing, Goethe-Universit\"at, D-60438 Frankfurt, Germany}

\author{Marko Horbatsch}
\email[]{marko@yorku.ca}
\affiliation{Department of Physics and Astronomy, York University, Toronto, Ontario M3J 1P3, Canada}

\author{Tom Kirchner}  
\email[]{tomk@yorku.ca}
\affiliation{Department of Physics and Astronomy, York University, Toronto, Ontario M3J 1P3, Canada}
\date{\today}
\begin{abstract}
Energy loss in collisions of charged projectiles with many-electron systems can be dealt with in time-dependent density functional theory
by invoking Ehrenfest's theorem for the time evolution of expectation values of observables.
We derive an exact expression for the evaluation of energy loss for systems described in a target reference frame, which is a functional
of the electron density.
Using an approximation scheme we then apply
the expression to antiproton-atom collisions at intermediate and high energies within the framework of the basis generator method.
The calculations are performed within the semiclassical approximation for the nuclear motion, and a straight-line trajectory
is employed. The energy loss is evaluated from an expectation value of the time derivative of the time-dependent projectile potential
and avoids the problem of identifying the excited and ionized many-electron contributions
in the many-electron wavefunction. 
There is also no need to invoke the independent-event model, since the calculations
are performed within the framework of the independent-electron mean-field model.
Detailed comparisons are provided for net ionization and total energy loss of antiprotons colliding with hydrogen,
helium, neon, carbon, nitrogen and oxygen. Reasonable agreement is found with the results from one-electron and two-electron
calculations for atomic hydrogen and helium, and with experiment in the latter case. For the $\bar p - \rm Ne$ system at intermediate collision energies 
we find discrepancies with previous
work that included only single-electron transitions. The sequence of results
for C, N, O, Ne allows one to paint a consistent picture which awaits experimental verification.
\end{abstract}
%
%

\maketitle
\section{Introduction}
\label{intro}
The description of ion-atom and ion-molecule collisions at intermediate energies has reached a high level of maturity. 
A number of methods have become available to deal with electronic excitations and ionization even for cases where the
target includes many electrons. In many instances comparison with experiment validates theoretical approaches. We will
focus in this work on attempts to approximate the time-dependent Schr\"odinger equation (TDSE) for a many-electron system
as it arises within the semiclassical approximation (classical treatment of nuclear motion), but it should be noted that there
is also a large body of works based on perturbative methods, including continuum distorted wave methods~\cite{Belkic2010}.

For biomedical applications, as well as those in material science an important role is played by the problem of energy deposition, namely how an impinging charged ion
transfers its initial kinetic energy to the matter. At the level of single collisions one refers to the notion of energy loss, which is mostly caused by 
the above-mentioned electronic processes. For positive ions the problem of capture complicates matters, since (partially or completely) neutralized projectiles
are created, and these interact with the medium in their own way in subsequent collisions, e.g., projectile ionization occurs, and ionization by neutral hydrogen
can be as effective as by proton impact. This complicates the situation for simulations of positive-ion passage through matter
which include secondary (and higher) collisions of particles produced in a first collision process.
For this reason we will restrict the discussion to antiproton
impact for which the only contributing processes to energy loss are electronic excitations to bound and continuum states of the target.
For molecular targets there are, of course, additional mechanisms such as dissociation or fragmentation.
The physics of antiproton-atom and antiproton-molecule collisions was reviewed, e.g., in Ref.~\cite{Kirchner_2011}.

One overview of the time-dependent density functional theory (TDDFT) approach to collisions involving many-electron targets that emphasizes projectile energy loss as a directly
accessible observable is Ref.~\cite{HJL2003}. The purpose of the present work is to expand on this idea of applying Ehrenfest's theorem in order to gain direct access
to energy loss via the electron density. 
The usefulness of this theorem in TDDFT was pointed out previously in Ref.~\cite{PhysRevLett.82.378}.

The traditional method of extracting this variable from ion-atom collision calculations relies on a projection of the evolved TDSE solution onto bound and continuum states
for a set of impact parameters, and to extract the transfer of projectile energy to the electron cloud by summing over electronic excitations. This is a difficult task which requires
an accurate propagation of continuum contributions.
The maturity of the field of extracting ionization information from TDSE calculations for antiproton-hydrogen collisions can be appreciated, e.g., from Fig.~10 in Ref.~\cite{PhysRevA.94.022703}, where state-of-the-art calculations for this single-electron atomic target are compared with experiment for total ionization as a function of collision energy.
This model was referred to as the (semiclassical) wavepacket convergent close coupling approach (WP-CCC), as compared to the purely Laguerre-basis
CCC approch, which is termed time-dependent CCC, and which was used for energy loss calculations~\cite{PhysRevA.92.022707}. 
A previous calculation based on a coupled-channel Lippmann-Schwinger scattering theory approach~\cite{Abdurakhmanov_2011}, labeled QM-CCC
provides very similar total ionization cross sections to WP-CCC, as demonstrated in Fig.~10 of Ref.~\cite{PhysRevA.94.022703}.
Single ionization from helium atoms at this level of theory was reported in Ref.~\cite{PhysRevA.96.022702}, and by a different method in Ref.~\cite{PhysRevA.90.052706} which
also allowed for the inclusion of double-ionization processes. Electron correlation effects on stopping power were dealt with in Ref.~\cite{PhysRevA.98.012707}.
A recent study using interesting methodology that can be implemented for molecular targets on the one hand confirmed the double ionization cross sections for He targets~\cite{PhysRevA.103.L030803},
but also pointed out how difficult it is to get a handle on extracting accurate double ionization cross sections.

The problem of stopping of antiprotons by atomic hydrogen and the rare gases was recently discussed in Ref.~\cite{PhysRevA.92.022707}.
The time-dependent CCC method was applied to helium both in a frozen-core (single active electron) and in a multi-configuration (MC) Hartree-Fock version,
and discrepancies were found at low and intermediate energies. The two-electron calculations are referred to as MC-CCC; for the single-active electron
calculations (hydrogen, neon) we will use the single-configuration SC-CCC as a name for the time-dependent CCC results.
For atomic hydrogen and helium atoms comparison is made with previous calculations, pointing out some discrepancies with atomic-orbital expansions and with distorted-wave results, 
but good overall agreement with
the calculations of L\"uhr and Saenz~\cite{PhysRevA.79.042901}. 
This work reports on molecular hydrogen calculations, as well, and these were
extended further in Ref.~\cite{PhysRevA.81.010701} with the use of two-electron wavefunctions to obtain single ionization data.
Further discussion of molecular hydrogen, and a model treatment of water molecule collisions  can be found in Refs.~\cite{PhysRevA.92.052711,BAILEY2016}.

In the context of material science stopping of antiprotons by excited Al clusters has been discussed within TDDFT in Ref.~\cite{KOVAL201356}. In this work the energy
loss of the antiproton is motivated by the acceleration experienced due to the time-dependent density response, which can be justified by an application of Ehrenfest's 
theorem \cite{HJL2003,PhysRevLett.82.378}, but in the end a straight-line trajectory is used. We also note that there is a literature on slowing of atoms in metals
and insulators treated by DFT approaches, and also more sophisticated methods with full correlation treatment~\cite{CUllrich, Correa2017}.

We would like to point out that within TDDFT one computes the projectile energy loss which is associated with a weighted sum over all $n-$fold electronic 
excitations, and is therefore a process based on net probabilities. For a discussion of energy loss and stopping power we refer the reader also to Ref.~\cite{PhysRevA.79.042901}.
The distinction between single-electron processes based loss and total energy loss (based on net ionization and excitation)  may be of lesser importance
for a system with singly charged projectiles and tightly bound electrons, such as helium, but one should be careful already with neon
and more so the heavier rare gases: for Ar, in particular, multiple ionization has been thoroughly investigated~\cite{PhysRevA.66.052719}.
For molecular targets with multiple weakly bound electrons the importance of 
multi-electron excitation is evident at lower and intermediate energies. This
has been discussed in the context of ionization of water molecules~\cite{PhysRevA.85.052704}, as well as molecules of biological 
interest~\cite{PhysRevA.101.062709,PhysRevA.99.062701,jorge2020multicharged}. For positively charged
projectiles the problem is amplified by the competition of capture and ionization~\cite{atoms8030059}.

The paper is structured as follows: in Section~\ref{sec:model} we derive the density functional for energy loss from the 
Ehrenfest expression for the time evolution of expectation values. In Section~\ref{sec:expt} we test the expression by comparing our results 
for atomic hydrogen against other calculations in the literature. This is followed by the helium target for which detailed two-electron calculations
were carried out by others and we demonstrate how the independent electron model fares in this case. For the targets involving $2s$ and $2p$ orbitals (the atoms 
neon, carbon, nitrogen,
and oxygen) we demonstrate how the present approach is capable of taking into account many-electron contributions to the energy loss.
Atomic units, characterized by $\hbar=m_e=e=4\pi\epsilon_0=1$, are used unless stated otherwise.

\section{Theory}
\label{sec:model}
\subsection{Definition of energy loss}
\label{sec:model1}

As outlined in the introduction most of the focus in the literature has been on single-electron or effective single-electron target descriptions,
except for some two-electron systems such as He~\cite{PhysRevA.92.022707} and molecular hydrogen~\cite{PhysRevLett.111.173201} 
for which one- and two-electron contributions were dealt with
explicitly.

Here we would first like to address the problem of energy loss in the context of $N$-electron systems, which can be obtained directly from the electronic density.
To be more specific, we concentrate on the electronic energy loss $S_e$, which is associated with the electronic excitations of the target system.
Momentum-transfer related contributions from projectile-nucleus target-nucleus scattering can play a role at low collision energies, and are ignored in the present work.
For a discussion of how one could deal with the nuclear contributions we refer the reader to the Appendix in Ref.~\cite{PhysRevA.92.022707}, 
but we caution that it is unclear whether these contributions should be simply treated in an additive way~\cite{Correa2017}.
Fig.~3 in Ref.~\cite{PhysRevA.92.022707} shows the relative contribution for the antiproton-helium system, and we conclude
that for the intermediate and high energies treated in the present work (we assume collision energies greater than $\rm 10 \ keV/amu$)
nuclear contributions to energy loss can be safely ignored~\cite{PhysRevLett.74.371}. 
As an aside we mention that the correlation between nuclear motion and ionization
was the object of experimental and theoretical studies in ion-atom collisions, and that these correlations can be investigated using classical trajectory
methods~\cite{PhysRevLett.63.147,Horbatsch_1989,DORNER200095}. 
Nevertheless, for the purpose of obtaining the energy loss cross section we can make the approximation $S \approx S_e$, and ignore the motion
of the target atom(s) during the collision. Therefore, we assume that a straight-line trajectory approach within the semiclassical approximation is
sufficient, and that the only explicit time dependence in the electron-nucleus interactions is associated with the projectile motion. 

The energy loss of the projectile during the collision with a many-electron target characterized by the electronic density $n(\vec r,t)$
(the diagonal part of the single-particle density matrix)
can be extracted from the total electronic energy given as a function of time, i.e., from
\begin{equation}
{\cal E} (t) = \langle \Psi(t) | H(t) | \Psi(t) \rangle \ .
\end{equation}
Here $\Psi(t)$ satisfies the many-electron TDSE, and the Hamiltonian is composed of kinetic energy, electron-nucleus attractions, as well as the electron-electron
repulsion (in the case of more than one electron present in the system), i.e.,
\begin{equation}
H(t)= {\hat T_e} +V_t + V_{ee} + V_p(t)\ .
\end{equation}
While working in the target reference frame and a system with $N$ electrons the explicitly time-dependent interactions can be expressed as
\begin{equation}
V_p(t) = \sum_{i=1}^N{\left(-\frac{Q_p}{r_p^{(i)}(t)} \right)} =  \sum_{i=1}^N{{\cal V}_p^{(i)}(t)} \ ,
\end{equation}
where ${\cal V}_p^{(i)}(t) \equiv {\cal V}_p({r_p}^{(i)}(t))$, and $Q_p$ is the projectile charge.
 
 The idea is to track the energy gain of the electrons during the collision, i.e.,
\begin{equation}
{\cal E}(t) - {\cal E}(t_i) = \int_{t_i}^t  \dot{\cal E}(t') \d t' \ .
\end{equation}
One needs to evaluate ${\cal E}(t_f)$ for a final time $t_f$ when the collision is over, and the transfer of energy has stopped. 
For a given nuclear
trajectory the quantity of interest is then the change in total electronic energy, i.e.,
${\cal E}(t_f) - {\cal E}(t_i)$. This electronic energy gain can be equated with projectile energy loss assuming negligible contributions
to the latter from direct transfer of energy between projectile and target nuclei. Thus, we will refer to energy loss ${\cal E}_L$ instead of
electronic energy gain.

Our interest is to treat the problem at the level of TDDFT, and express ${\cal E}_L (t)$ in terms of the density $n(\vec r,t)$~\cite{CUllrich}.
This approach avoids the need to solve the many-electron TDSE, while the time-dependent Kohn-Sham (KS) equations provide orbitals to represent
the time evolution of the density. Explicitly, we can write
\begin{equation}
\dot {\cal E}_L(t)=\langle \Psi(t)|\partial_t H(t) |\Psi(t) \rangle = \langle \Psi(t)|\sum_{i=1}^N{\partial_t{\cal V}_p^{(i)}} |\Psi(t) \rangle  = \int{n({\vec r},t){\dot {\cal V}_p(\vec r,t)}  \d \vec r} \ ,
\end{equation}
where in the final expression the explicit time derivative of the single-particle operator ${\cal V}_p(t)$ appears.
This potential energy and its time derivative are written explicitly as
\begin{eqnarray}
  {\cal V}_p(\vec r,t) &=& -\frac{Q_p}{r_p(t)} = -\frac{Q_p}{\mid \vec r - \vec R(t)\mid} \nonumber \\
  \dot{\cal V}_p(\vec r,t) &=& \dot{\vec R} \cdot \nabla_R {\cal V}_p(\vec r,t)  \ ,
  \label{eq:Vp}
\end{eqnarray}
where $\vec R(t)$ is the position vector of the projectile as viewed from the target.

The time evolution of the projectile energy loss for a  single collision then follows as
\begin{equation}
  {\cal E}_L(t) =  {\cal E}(t) - {\cal E}(t_i) = \int_{t_i}^t  \int  n(\vec r,t') \dot{\cal V}_p (\vec r, t') \d \vec r \d t'  
  \equiv \int_{t_i}^{t}{\langle \dot{\cal V}_p ( t') \rangle \d t'} \; ,
   \label{eq:EL}
\end{equation}
and obvioulsy before the collision we have ${\cal E}_L(t_i) = 0$.
The electronic contribution to the projectile energy loss, $S_e$,  is one of few observables which can be represented as an exact functional of $n(\vec r,t)$.

The main idea is now to solve for the energy loss function in parallel with the  Kohn-Sham equations based on a time-stepping scheme
for the expectation value  $\langle  \dot{\cal V}_p (t) \rangle $:
\begin{equation}
  {\cal E}_L(t+\Delta t) =  {\cal E}_L(t) +  \Delta t  \int  n(\vec r,t) \dot{\cal V}_p (\vec r, t) \d \vec r   \  .
  \label{eq:IntGl}
\end{equation}

Some remarks for the evaluation of the required expectation value are in order. For a straight-line trajectory of the projectile,
while working in an inertial reference frame
of the target system one can always specify the explicitly time-dependent potential contribution as indicated in eq.~(\ref{eq:Vp}).
Assuming the motion of the projectile to be parallel to the $z-$axis, we can express the nuclear trajectory specified by impact
parameter $b$ and impact velocity $v$ as $\vec R(t) = (b,0,v t)$, and we thus obtain for the time derivative of the potential
\begin{equation}
   \dot{\cal V}_p(\vec r,t) =   -Q_p  \;\frac{\dot{\vec R} \cdot \vec r_p(t)}{r_p(t)^3} =  -Q_p v \; \frac{z_p(t)}{r_p(t)^3}  \ .
    \label{eq:Vdot}
\end{equation}
It is at this point, namely the assumption of a straight-line trajectory, that we neglect nuclear contributions to energy loss.

We observe that the energy loss functional vanishes for a constant density $n(\vec r)$ which is symmetric in $z$, due to a symmetry property of $\dot{\cal V}_p$
under $t \to -t$, and we find
\begin{eqnarray}
\int_{t_i}^{t_f}  \int  n(\vec r,t_i) \dot{\cal V}_p (\vec r, t') \d \vec r \d t' \nonumber \\
                     & = &  \int  n(\vec r,t_i) \int_{t_i}^{t_f}  \dot{\cal V}_p (\vec r, t') \d t'  \d \vec r = 0 \ ,
\end{eqnarray}
provided the boundaries $t_i,t_f$ are chosen symmetrically around the time of closest approach (chosen to be $t=0$).
The symmetry constraint for the initial-state density is only necessary for the vicinity of the closest approach. If one chooses the 
 boundaries $t_i,t_f$ such that the internuclear separation is much larger than the distance scale where the initial-state density is non-zero,
 the subtraction scheme is practically valid for all inital-state densities $n(\vec r)$.

This can be used for a computationally convenient subtraction scheme and results in the alternative evaluation of loss according to
\begin{equation}
  \tilde {\cal E}_L(t_f) =   \int_{t_i}^{t_f}  \int  [n(\vec r,t)-n(\vec r,t_i)] \; \dot{\cal V}_p (\vec r, t) \d \vec r \d t
    \equiv \int_{t_i}^{t_f}{\langle \dot{ {\widetilde{ \cal{V}}}}_p ( t) \rangle \d t} \, . 
   \label{eq:drho} 
\end{equation}
While it is true that $\tilde {\cal E}_L(t_f) \equiv {\cal E}_L(t_f)$ in the limit of $t_f=-t_i$ being large, the time evolution of the two quantities is very different, as will be shown in Section~\ref{sec:prelim} for some examples.

The usual summation over impact parameters is required in order to calculate the projectile energy loss at a given impact velocity.
Using the impact energy $E=m_p v^2/2$, where $m_p$ is the projectile mass, one integrates  ${\cal E}_L(b,E) \equiv {\cal E}_L(t_f)$  over $b$,
 to arrive at the energy loss for a given projectile energy  $E$, viz.,
\begin{equation}
  S_e (E) = 2\pi \int _0^\infty{{\cal E}_L(b,E) b \d b} \; .
\end{equation}
This quantity is called energy-loss cross section, or stopping cross section, and is usually given in units of $\rm (eV cm^2)$ or (\AA${}^2$eV). 
The so-called electronic stopping power is defined as
\begin{equation}
  -\frac{\d E}{\d x} = n_A S_e(E)
\label{eq:eloss}
\end{equation}
where $n_A$ is the number density of the medium, and small fluctuations (straggling) are assumed~\cite{PSigmund,PhysRevA.92.022707}. 

\subsection{Calculation of energy loss in  a subspace of Hilbert space}
\label{sec:model2}
We now proceed with comments on how to implement  projectile energy loss in the framework of a projector formalism
as it appears, e.g., in  a finite-basis representation of the TDSE, and in the basis generator method (BGM) in particular. 
A review of the BGM can be found in Ref.~\cite{hjl18}, while detailed results for ionization
of atoms which are the building blocks of many biological molecules can be found in Refs.~\cite{PhysRevA.101.062709,hjl19}.

Let $\cal{P}$ be a subspace of Hilbert space, $\hat P$ the appropriate projector, and $\ket{\Psi_P}$ a solution of the TDSE
{\it within} the subspace $\cal{P}$:
\begin{equation} 
\hat P \idt \ket{\Psi_P}= \hat P \hat H  \ket{\Psi_P} = \hat P \hat H \hat P  \ket{\Psi_P} \equiv \hat H_{PP}  \ket{\Psi_P} \ .
\end{equation}
Note that $\ket{\Psi_P(t)}$ is not the exact  TDSE solution $|\Psi(t) \rangle$  projected onto $\cal{P}$, but represents
an approximate TDSE solution.
The time derivative of the energy expectation value can be evaluated as
\begin{eqnarray}
 \dot{\cal E}^P &\equiv&  \frac{\d}{\d t} \bra{\Psi _P(t)} \hat H(t) \ket{\Psi_P(t)} \nonumber \\
                    &=& \bra{\Psi_P (t)} \partial _t \hat H_{PP}(t) -\i [\hat H(t), \hat H_{PP}(t)]\ket{\Psi_P(t)} \nonumber \\
                    &=& \bra{\Psi_P (t)} \partial _t \hat H_{PP}(t) \ket{\Psi_P(t)} \ .
\end{eqnarray}
Here we used the property that the expectation value of the commutator vanishes within the subspace $\cal{P}$.

We should distinguish two possible scenarios, namely whether the projector $\hat P$ is explicitly time dependent or not.
Explicit time dependence will certainly occur within the two-center BGM approach for positive ion impact collisions
on account of evolving projectile states, as well
as the application of the projectile hierarchy associated with the $W_p$ operator (cf. eqs.~(6,7) in Ref.~\cite{hjl18}).
This case is complicated and requires further thought. The other case, for which there is no explicit time dependence in the
projector ($\partial_t \hat P = 0)$ is more straightforward, and can be realized, e.g., within the target ($W_t$) hierarchy in one-center 
BGM~\cite{PhysRevA.67.062711,Kirchner_2002}, which is suitable for antiproton impact.
This case also applies to TDSE calculations using spatial discretization (whether using a grid, finite elements, or a pseudospectral representation).

We may write in this case

\begin{eqnarray}
 \bra{\Psi_P (t)} \partial _t \hat H_{PP}(t) \ket{\Psi_P(t)} &=& \bra{\Psi_P (t)} \partial _t (\hat P\hat H(t)\hat P) \ket{\Psi_P(t)} \nonumber \\
                                         &=& \bra{\Psi_P (t)} \partial _t \hat H(t) \ket{\Psi_P(t)}\nonumber \\
                                         &=& \bra{\Psi_P (t)} \partial _t \hat {\cal V}_p(t) \ket{\Psi_P(t)} \ , 
\end{eqnarray}
i.e., Ehrenfest's theorem remains unchanged and we may proceed to
 consider a basis representation of eq.~(\ref{eq:IntGl}) within a stationary BGM basis~\cite{PhysRevA.67.062711}.
A KS orbital $\psi_{\gamma} (\vec r,t)$ is represented as
\begin{equation}
  \psi_{\gamma} (\vec r,t) = \sum_{j=1}^N \sum_{J=0}^{M} a_{jJ}^\gamma (t) \chi_j^J(\vec r)
\end{equation}
with the following definitions: the basis function $\chi_j^J(\vec r) \equiv \langle \vec r | j J \rangle$ is obtained by $J$-fold
application of the Yukawa regularized target interaction $W_t$  on the generating atomic orbital (AO) basis, i.e.,
\begin{equation}
  \ket{jJ} = W_t^J \ket{j0}  \ .
\end{equation}
The latter is specified by
\begin{eqnarray}
  {\hat h}_t  \ket{j0} &=& \varepsilon _j \ket{j0}  \ 
\end{eqnarray}
with the single-particle target Hamiltonian
\begin{eqnarray}
  {\hat h}_t  = - \frac{1}{2} \nabla^2  + v_t 
\end{eqnarray}
in which $v_T$ is the ground-state potential at the level of exchange-only density functional theory.
The many-electron system is thereby treated at the level of the independent-particle model~\cite{PhysRevA.61.012705,PhysRevA.62.042704}.
The single-particle density of the system is represented by the KS orbitals and can now be expressed as
\begin{eqnarray}
  n(\vec r,t) &=& \sum_\gamma \scp{\vec r}{\psi_\gamma} \scp{\psi_\gamma}{\vec r}\nonumber\\
                  &=& \sum_\gamma \sum_{jJ}\sum_{kK} a_{jJ}^\gamma (t) a_{kK}^{\gamma *} (t) \scp{\vec r}{jJ} \scp{kK}{\vec r}
\end{eqnarray}
Insertion into eq.~(\ref{eq:IntGl}) yields the expression
\begin{eqnarray}
   {\cal E}(t+\Delta t) =  {\cal E}(t) + \Delta t  \sum_\gamma \sum_{jJ}\sum_{kK} a_{jJ}^\gamma (t) a_{kK}^{\gamma *} (t) M_{kj}^{KJ}
\end{eqnarray}
with
\begin{eqnarray}
    M_{kj}^{KJ}          =  \bra{kK} \, \dot{\cal V}_p \, \ket{jJ}
                                = -Q_p v \; \bra{kK} \, \frac{z_p(t)}{r_p(t)^3} \, \ket{jJ} \; .
\end{eqnarray}

\section{Results}
\label{sec:expt}

\subsection{Demonstration of convergence}
\label{sec:prelim}

We begin our discussion of results by demonstrating the principle of how the calculation works within the one-center BGM approach using a target $W_t$
hierarchy for antiproton-hydrogen collisions. The aim is to show the effectiveness of the subtraction scheme eq.~(\ref{eq:drho}) in comparison with the direct evaluation of eq.~(\ref{eq:EL}).

\begin{figure}
\begin{center}$
\begin{array}{cc}
\resizebox{0.5\textwidth}{!}{\includegraphics{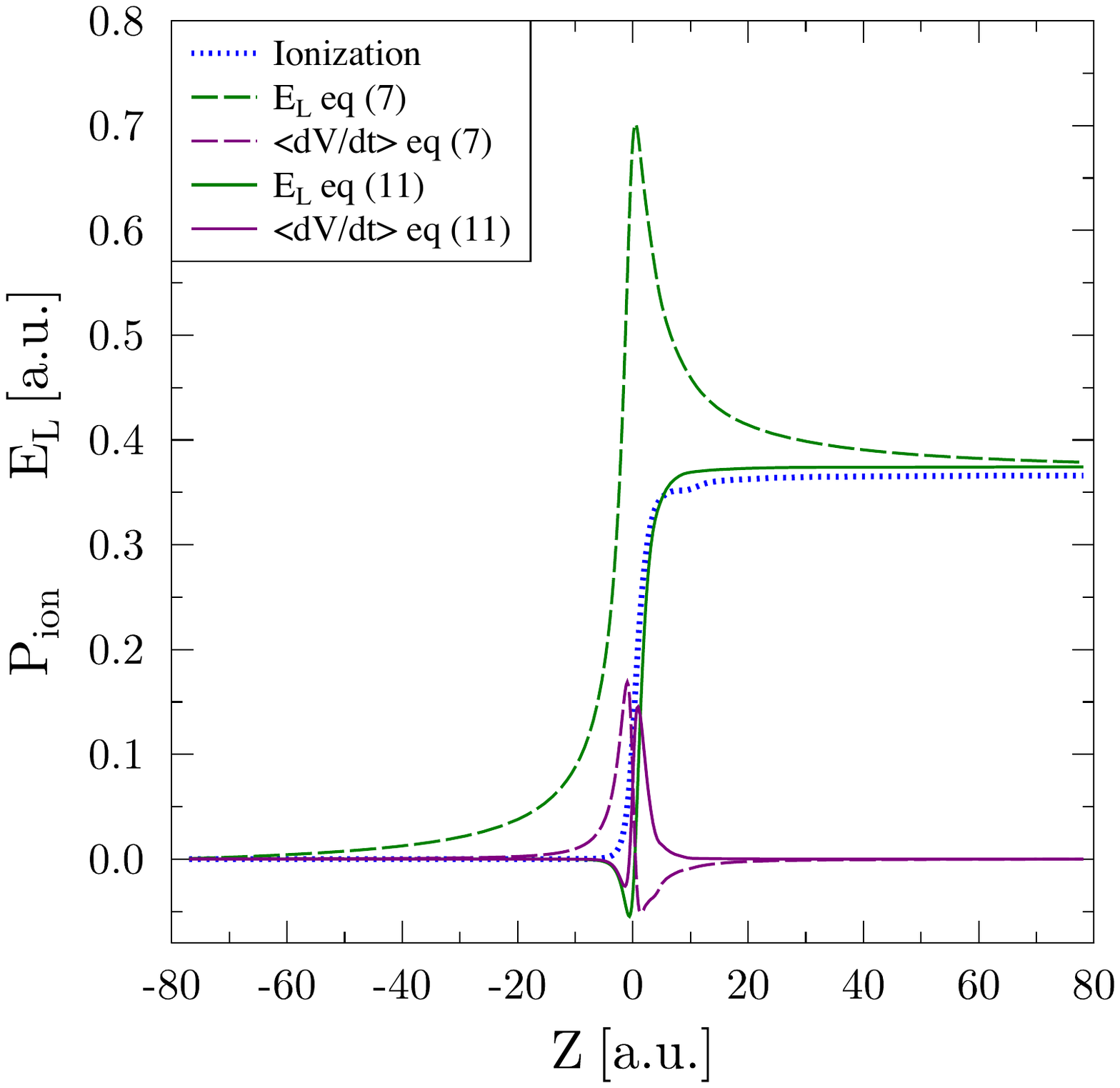}}{\hspace{-0.5 truecm}}&
\resizebox{0.5\textwidth}{!}{\includegraphics{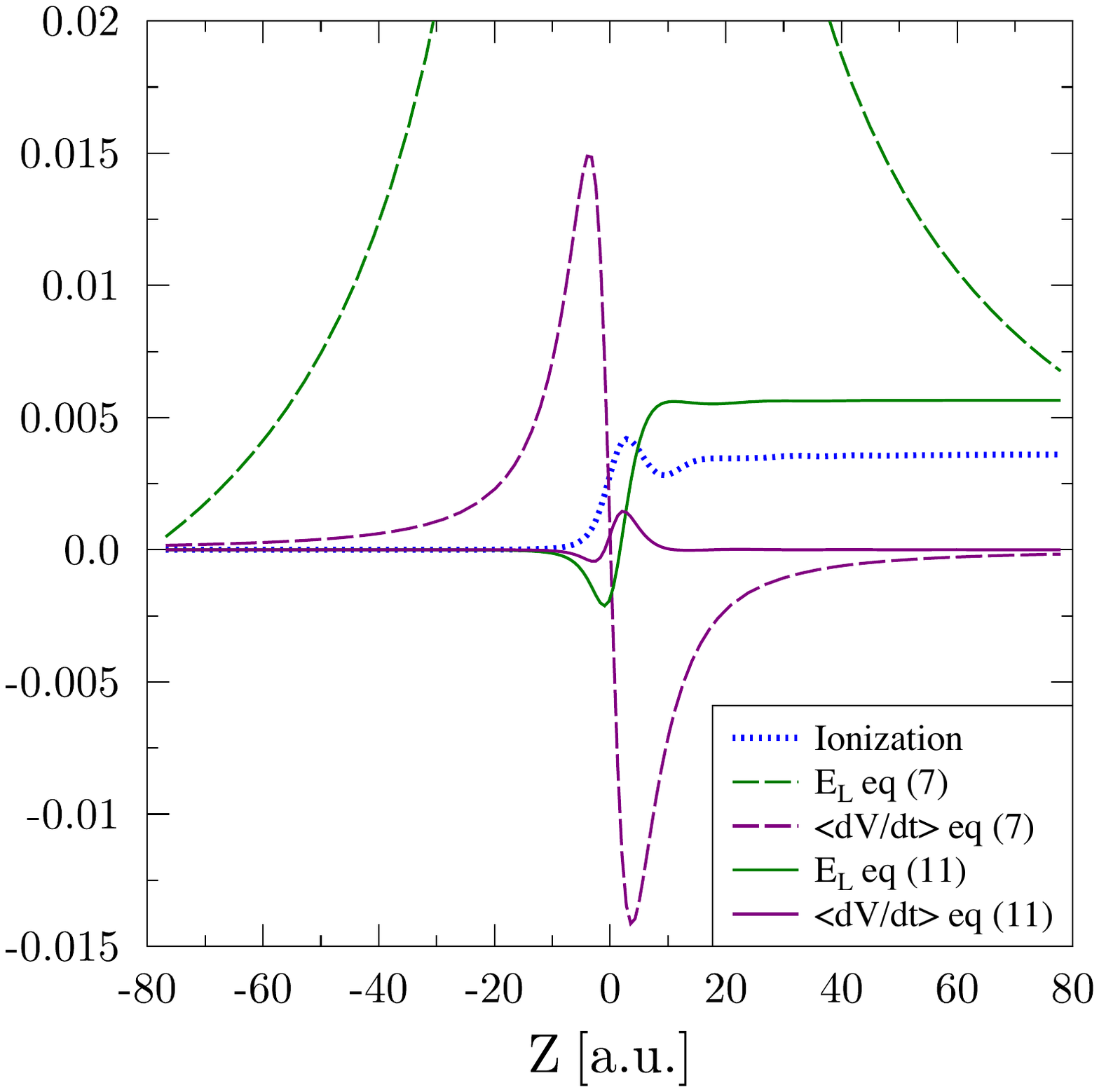}}

\end{array}$
\caption{%
Time evolution of some variables during an antiproton-hydrogen collision at 25 $\rm keV/amu$ with impact parameters $b=1$
in the left panel, and $b=5$ in the right panel.
The BGM ionization probability is shown as a dotted blue line.
The expectation values of the rate of change of the projectile potential energy $\langle  \dot {\cal V}_P(t)  \rangle $ or $\langle  \dot { {\widetilde{ \cal{V}}}}_P(t)  \rangle $, 
as they appear in eqs.~(\ref{eq:EL}) and~(\ref{eq:drho}) respectively are shown as dashed and solid purple lines.
The time integrals themselves, from eqs.~(\ref{eq:EL}) and~(\ref{eq:drho}) which both yield the
projectile energy  loss for the given value of $b$ when $t_f=-t_i$ is sufficiently large are shown as green dashed and solid lines. 
The time evolution is expressed through the
nuclear coordinate $Z(t) = v t$ given on the horizontal axis, where $t=0$ corresponds to closest approach, and $t_f=-t_i=80$.
While the probabilities are dimensionless, the other variables are given in atomic units.
}
\label{fig:Fig1}
\end{center}
\end{figure}

In Fig.~\ref{fig:Fig1} the time evolution of the relevant quantities is shown for an impact velocity of $v=1$ ($E_p=25 \ \rm keV$) and two impact parameters, namely $b=1$ which contributes significantly to the total energy loss, and a distant impact parameter, $b=5$ to illustrate the usefulness of the subtracted energy loss evaluation according to eq.~(\ref{eq:drho}). 
We notice an overshoot in the time evolution of ${\cal E}_L(t)$ according to eq.~(\ref{eq:EL}) in the vicinity of the closest approach for $b=1$, and over a large range for $b=5$.
The primary reason is the Stark shifting of the initial state as the projectile approaches; while the projectile recedes after the collision the Stark shifting changes due to
the de-population of the initial state. This is responsible for the asymmetry in the time evolution of ${\cal E}_L(t)$ calculated according to eq.~(\ref{eq:EL}) (dashed green lines).

The energy loss ${\cal E}_L(t)$ is caused by population of discrete excitations, as well as transfer of population to the continuum. Some of these excitations are temporary, i.e.,
some population is transferred back to the ground state, and this causes  some oscillations. The trace of $\tilde {\cal E}_L(t)$ displays a small dip before closest approach:
this is caused by the depopulation of the initial state, and thus the subtraction leads to a swing towards negative values.
Obviously, the time evolution of ${\cal E}_L(t)$ (or of $\tilde {\cal E}_L(t)$) is not an observable, and only the value for sufficiently large $t_f$ matters. The same is true for
the ionization probability as a function of time.

Two advantages can be observed for the evaluation of loss according to eq.~(\ref{eq:drho}) vs eq.~(\ref{eq:EL}): the overshoot feature
at the closest approach disappears, but more importantly, the time variation is limited to a smaller time interval around the closest approach.
Evaluation according to eq.~(\ref{eq:drho}) is absolutely necessary to reach the required accuracy at large impact parameter values. The calculation
of $\tilde {\cal E}_L(t_f)$ can be obtained with a substantially reduced time interval as compared to that of ${\cal E}_L(t_f)$.
The comparison of evaluating $\tilde {\cal E}_L(t_f)$  versus $ {\cal E}_L(t_f)$ is reflected also in the traces of the expectation value of the 
rate of change of the projectile potential energy, i.e., 
$\langle \dot{ {\widetilde{ \cal{V}}}}_p(t) \rangle$  (solid line),
showing rapid convergence as opposed to $\langle \dot{ \cal V}_p(t) \rangle$ (dashed line).

\begin{figure}
\begin{center}
\resizebox{0.7\textwidth}{!}{\includegraphics{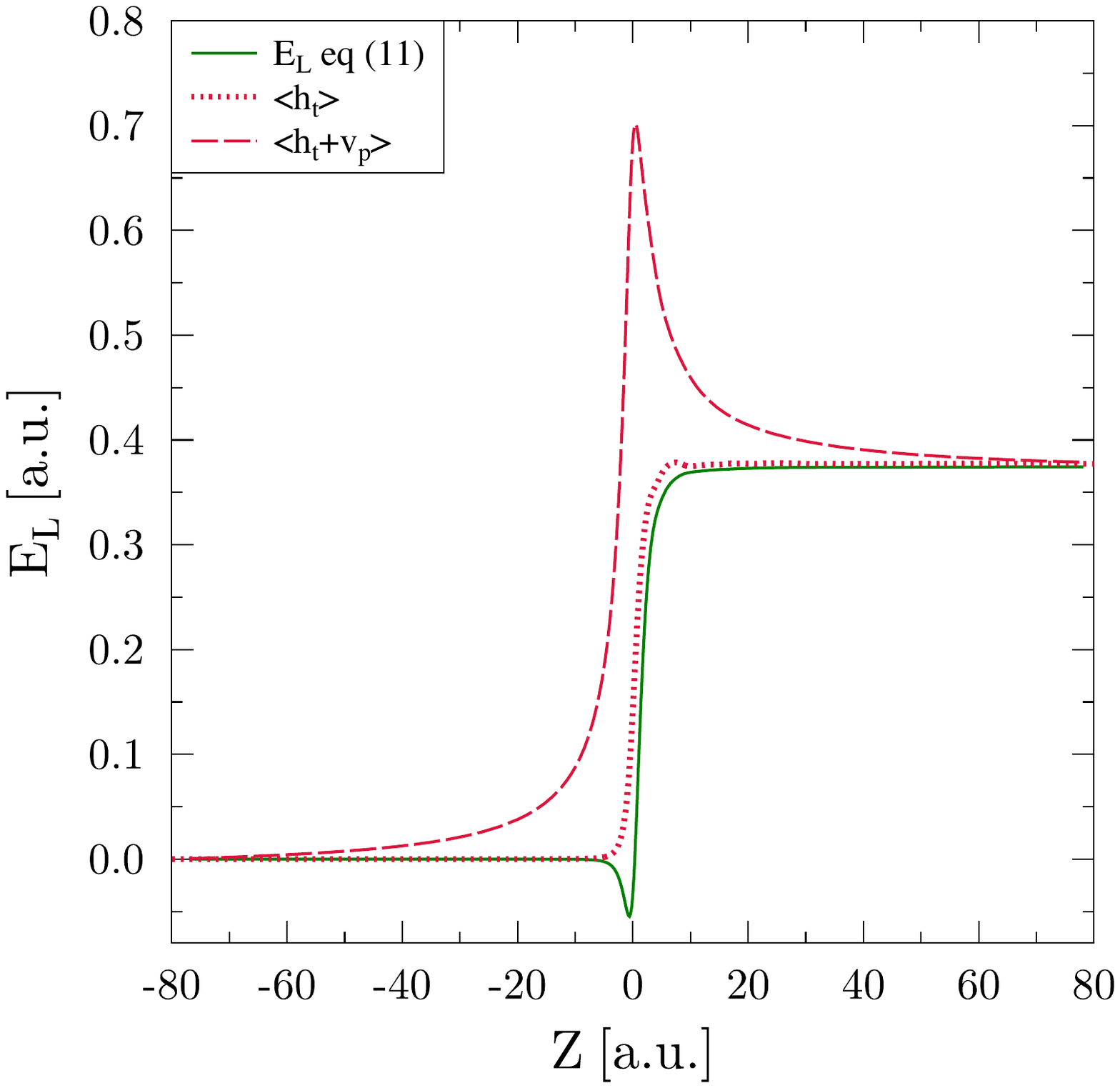}}

\caption{Comparison of energy loss for $b=1$, $v=1$ calculations according to eq.~(\ref{eq:drho}) (shown as a solid green line)
with the time evolution of the 
expectation value of the target Hamiltonian $\langle \psi(t)| {\hat h_t} |\psi(t) \rangle - \epsilon_i$ (dotted red line), where $\epsilon_i$ is the eigenvalue of the initial state,
and with the expectation value of the full Hamiltonian $\langle \psi(t)| {\hat h_t} +{\cal V}_P(t) |\psi(t) \rangle -(\epsilon_i+Q_p/R(t_i)) $ (dashed red line).
The energy loss is given in atomic units.}
   \label{fig:Fig2}
\end{center}
\end{figure}

Another interesting comparison can be made for the evaluation of energy loss. In Fig.~\ref{fig:Fig2} we compare the Ehrenfest theorem based evaluation of $\tilde {\cal E}_L(t)$
with the traditional way of extracting loss via the expectation value of the target Hamiltonian, i.e., $\langle \psi(t)| {\hat h_t} |\psi(t) \rangle -\epsilon_i $,
where $\epsilon_i $ is the eigenenergy of the initial state.
The point we are making here is that the density functional based evaluation of $\tilde {\cal E}_L(t_f)$ is competitive with the traditional calculation. While this may
not be essential for a single-electron system, using $\tilde {\cal E}_L(t)$ does represent a definite advantage for many-electron targets.
The result shown as the dashed red line is based on the expectation value of
the full Hamiltonian (which includes the Stark perturbation) 
 $\langle \psi(t)| {\hat h_t} +{\cal V}_P(t) |\psi(t) \rangle - (\epsilon_i+Q_p/R(t_i)) $,
 and it results in a curve completely equivalent to  ${\cal E}_L(t_f)$ (shown as the dashed green line
in the left panel of Fig.~\ref{fig:Fig1}).

\begin{figure}
\begin{center}$
\begin{array}{ccc}
\resizebox{0.45\textwidth}{!}{\includegraphics{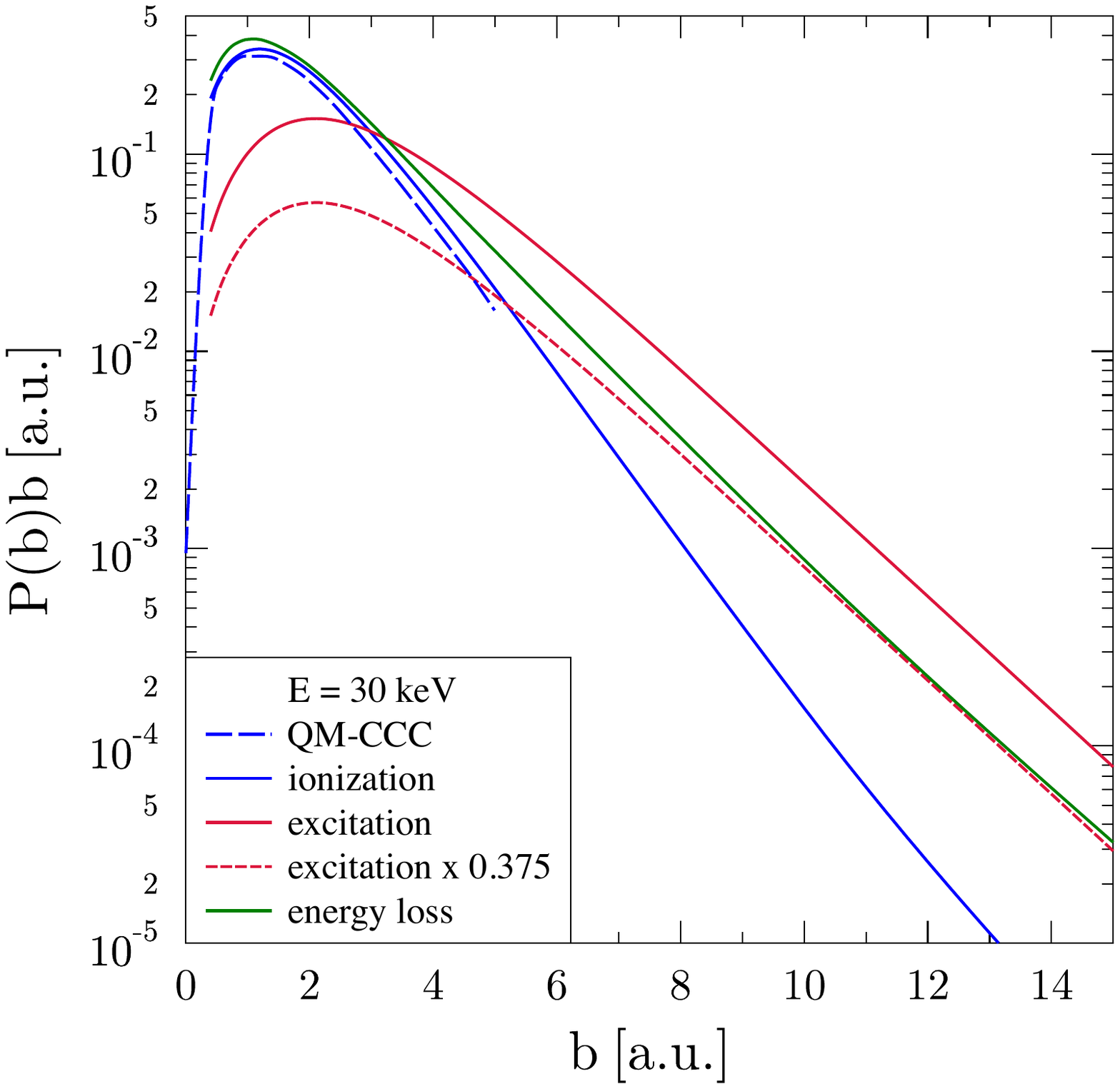}}{\hspace{-2.5 truecm}}&
\resizebox{0.45\textwidth}{!}{\includegraphics{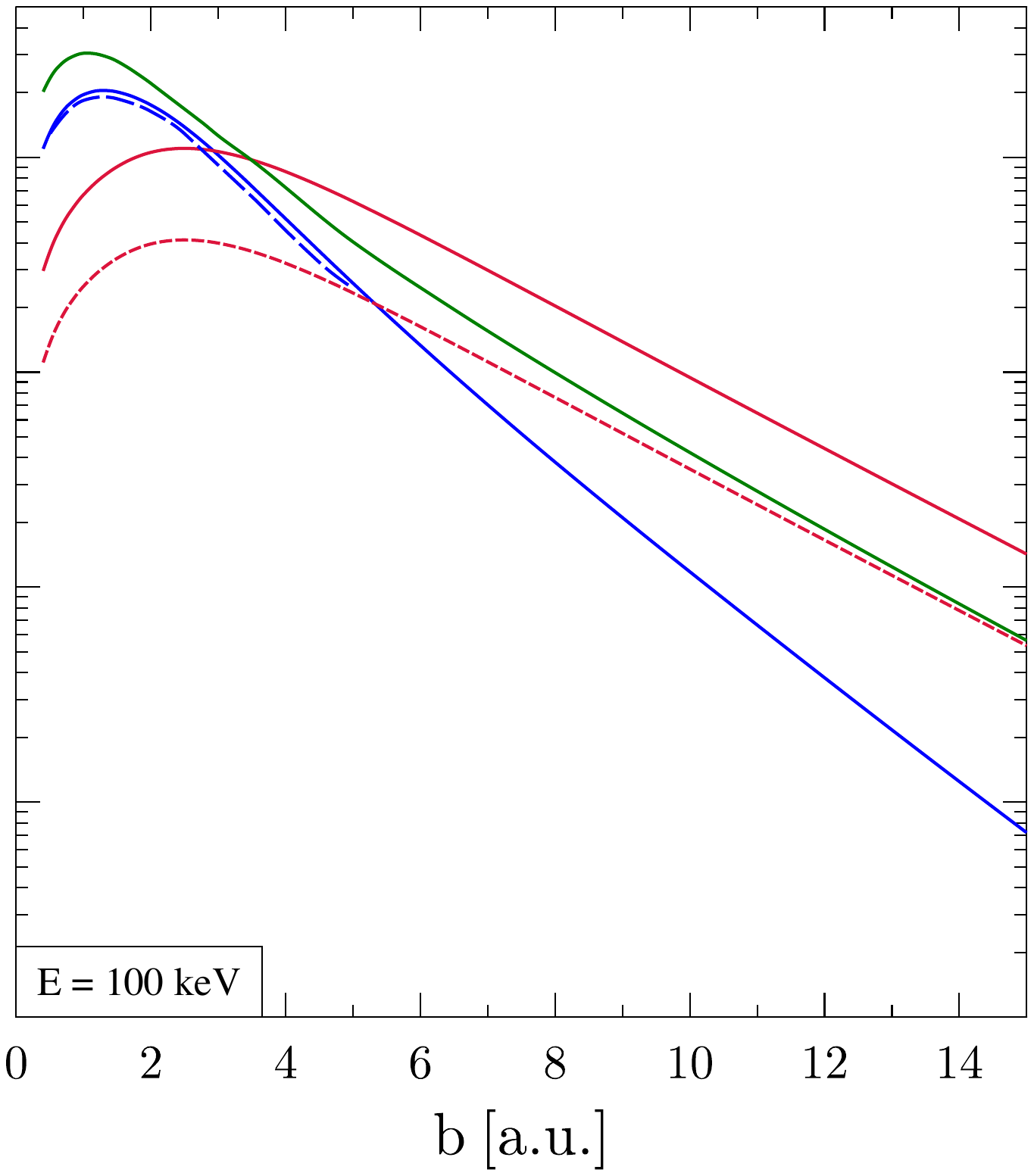}}{\hspace{-2.5 truecm}}&
\resizebox{0.45\textwidth}{!}{\includegraphics{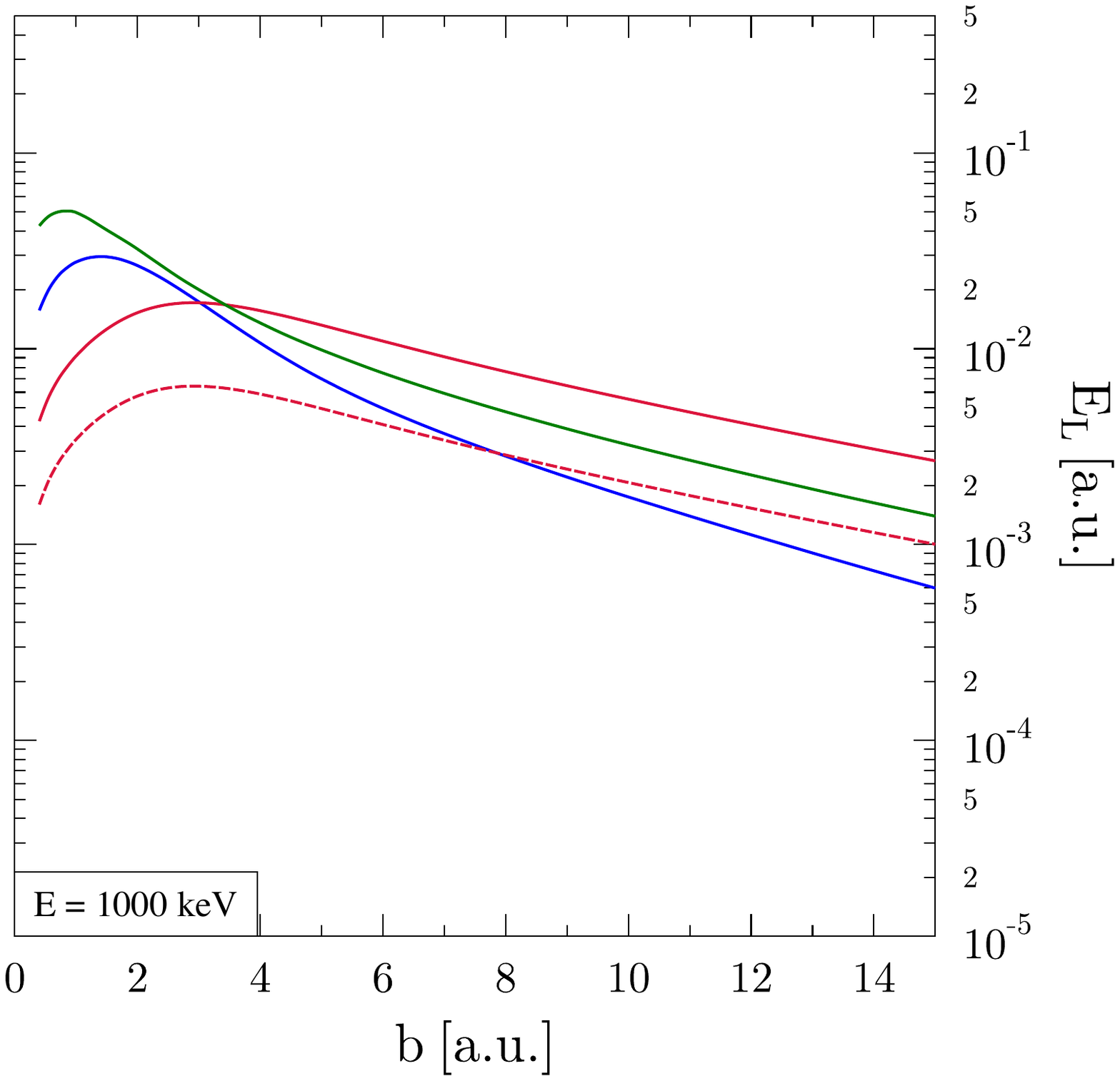}}
\end{array}$
{\vspace{-1.5 truecm}}
\caption{Impact parameter dependence of ionization and discrete excitation probabilities, as well as energy loss
for antiproton collisions with atomic hydrogen at collision energies of $30 \ \rm keV$, $100 \ \rm keV$, and $1000 \ \rm keV$ are shown in the
left, middle, and right panels respectively. 
The probabilities are weighted with $b$ to show how they contribute to the total
cross sections. 
The blue solid line shows the BGM ionization probability, the blue dashed lines for 30 and 100 keV are the QM-CCC results~\cite{Abdurakhmanov_2011}.
The red solid lines show the BGM probabilities for total discrete excitations (summing over shells $n=2,3,4$), while the red dashed lines 
show this excitation probability times $0.375=1/8-1/2$. Assuming dominant excitation to $n=2$ this corresponds to energy loss at large $b$.
The $b$-weighted energy loss (in atomic units) is represented by green solid lines.
}
   \label{fig:Fig3}
\end{center}
\end{figure}

In Fig.~\ref{fig:Fig3} we compare $b$-weighted probabilities for ionization with the QM-CCC results of Ref.~\cite{Abdurakhmanov_2011}.
We notice overall good agreement between the calculations, and find that the present BGM results are higher by less than $10 \%$ starting in the
vicinity of the maximum, which occurs somewhat to the right of $b=1$ for impact energies of 30 and 100 keV. 
Ionization provides the main contribution to energy loss, which can be seen
to peak in a similar $b$-range. The energy loss probability initially follows the ionization curve, 
except for the larger impact parameters where discrete excitations make a sizeable
contribution to loss.

The red dashed curve shows the discrete excitation probability (summed over $n=2,3,4$) multiplied by the energy for excitation to $n=2$.
This probability merges with the energy loss probability at large impact parameters. Excitation to an $n=2$ Stark state (linear combination of $2s$ and $2p$) 
occurs during the collision.
For very high energies (right panel) the merging occurs at $b>15$, i.e., outside the shown range.

Note that the logarithmic presentation may be deceiving in the following regard: the energy loss is, in fact dominated by ionization in the sense
that the area under the green curve comes mostly from small and intermediate values of $b$ where ionization dominates. For distant collisions 
ionization is less effective than discrete excitation, but the probabilities are small, and the weighting with the excitation energy further suppresses
the contribution to energy loss. Nevertheless, it is interesting to observe the turnover in the energy loss traces for 100 and 1000 keV, where the
slope changes at large values of $b$.

The present calculations for hydrogen targets were tested against basis set convergence within the method. From these tests
we derive confidence of obtaining ionization cross sections with an uncertainty of about $5 \%$.

\subsection{Collisions of antiprotons with atoms}
\label{sec:expt1}
In this section we present calculations of  projectile energy loss for antiproton collisions with a number of atomic targets which
are relevant for independent atom model calculations for molecules of biological interest. The selected atomic targets are hydrogen (H), then the rare gases
He and Ne, and finally some atoms that are constituents of many molecules, such as  carbon (C), nitrogen (N), and oxygen (O).
For the target atoms with more than one electron
calculations were carried out with frozen atomic target potential during the collision (called no-response), 
and also within a model where the mean-field potential is motivated by TDDFT (called dynamic response model)~\cite{PhysRevA.62.042704}.

\subsubsection{Hydrogen}
\label{ssec:h}

\begin{figure}
\begin{center}$
\begin{array}{cc}
\resizebox{0.5\textwidth}{!}{\includegraphics{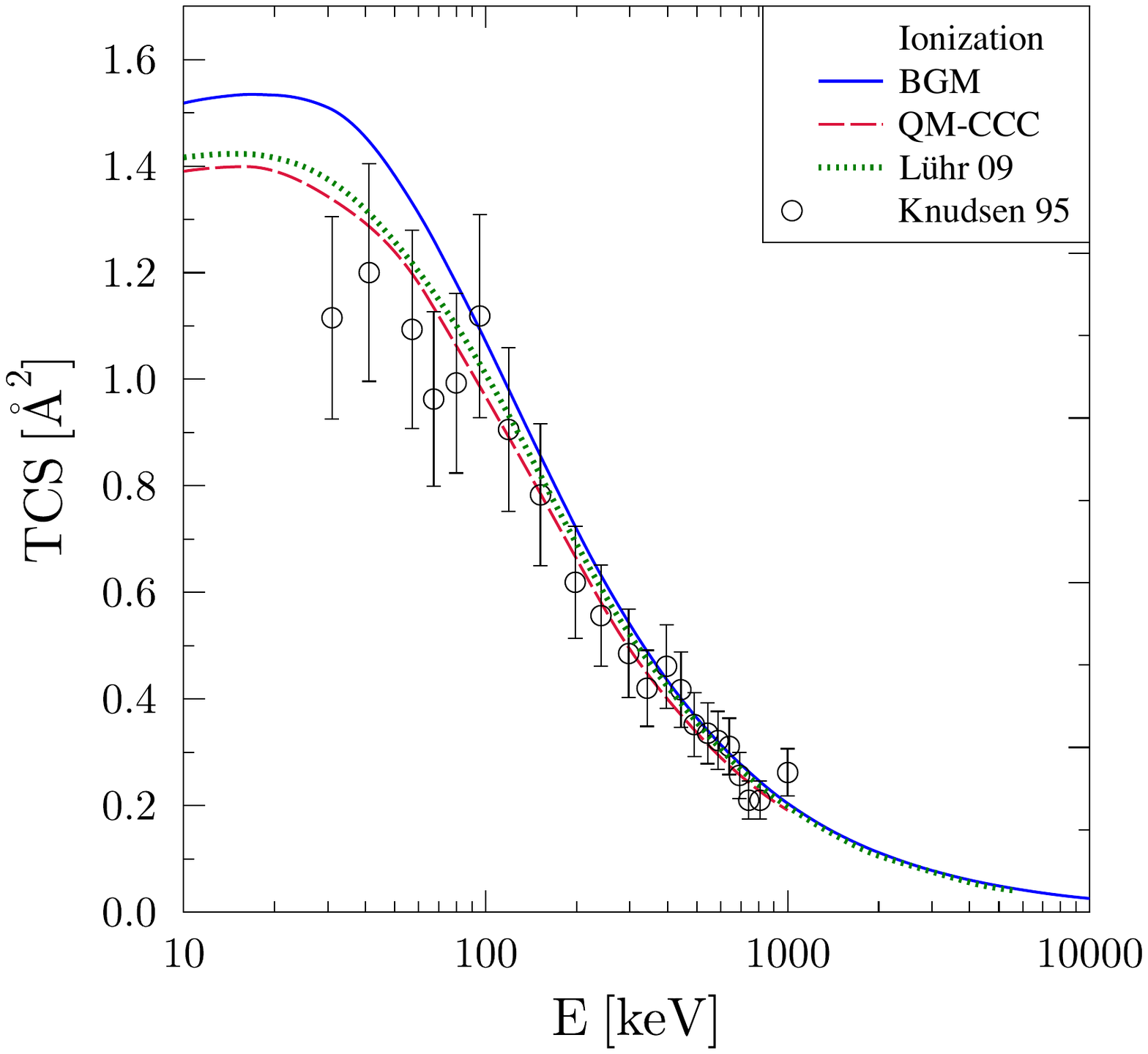}}{\hspace{-0.5 truecm}}&
\resizebox{0.5\textwidth}{!}{\includegraphics{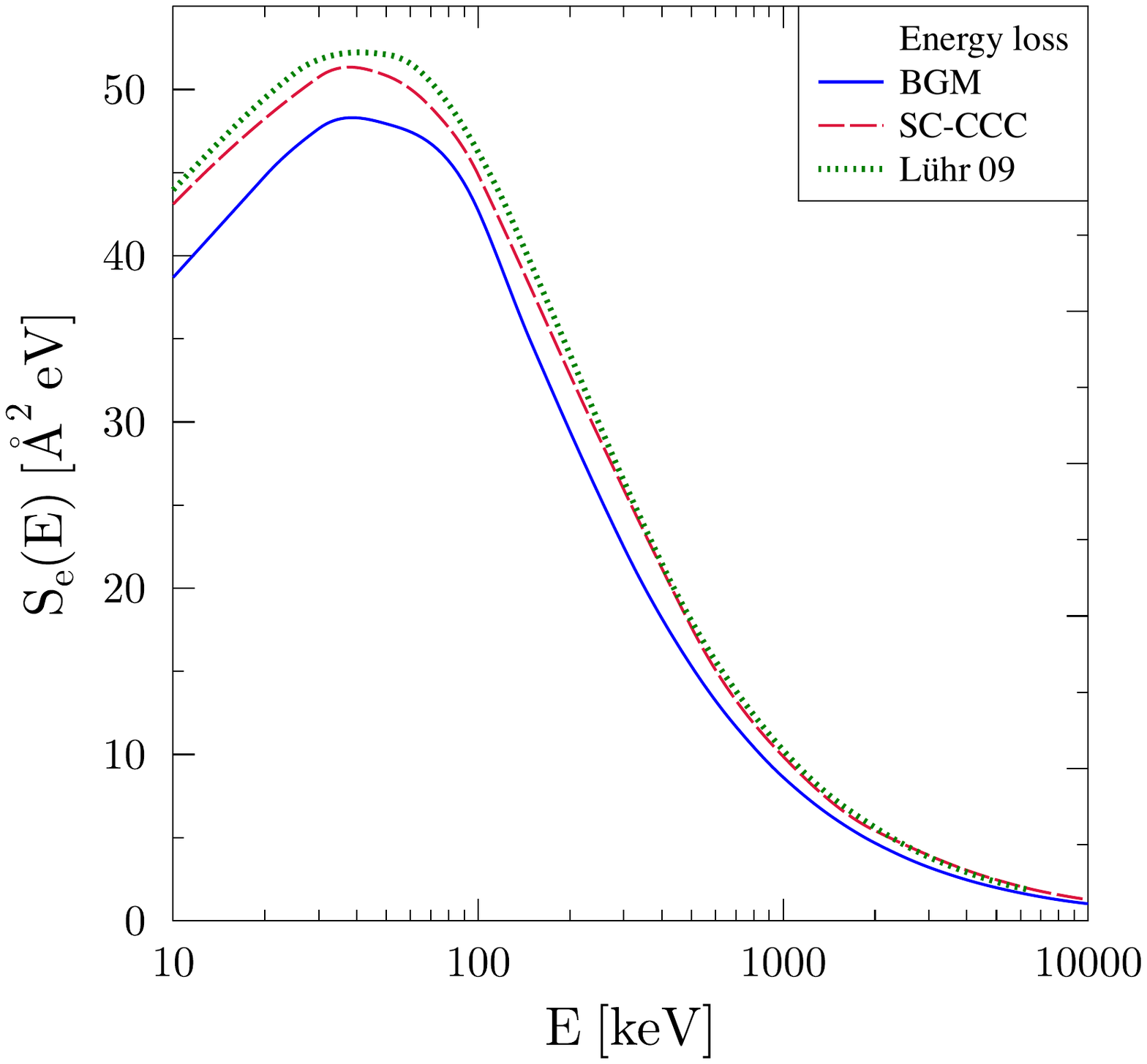}}
\end{array}$
{\vspace{-1.5 truecm}}
\caption{Energy dependence of the ionization cross section (left panel), and the energy loss cross section (right panel)
for antiproton collisions with atomic hydrogen. 
The BGM ionization cross section is shown as a blue solid line.
The dotted green line is from Ref.~\cite{PhysRevA.79.042901}, while the dashed red line is the QM-CCC result of
Ref.~\cite{Abdurakhmanov_2011}.  
The experimental ionization data are from Ref.~\cite{PhysRevLett.74.4627}.
The right panel shows the BGM energy loss (solid blue line),
the SC-CCC result~\cite{PhysRevA.92.022707} as a dashed red line, and the result from Ref.~\cite{PhysRevA.79.042901} as a dotted green line.
}
\label{fig:Fig4}
\end{center}
\end{figure}

We begin with total cross sections for atomic hydrogen targets.
The energy dependence of the total ionization cross section is compared in the left panel of Fig.~\ref{fig:Fig4} with experiment~\cite{PhysRevLett.74.4627}, 
and with two previous calculations reported in Refs.~\cite{PhysRevA.79.042901,Abdurakhmanov_2011}. At impact energies $E>100 \ \rm keV$ they all agree very well.
At lower energies the BGM results are higher than the results of L\"uhr and Saenz~\cite{PhysRevA.79.042901}, as well as the QM-CCC results of
Abdurakhmanov {\it et al.}~\cite{Abdurakhmanov_2011}.
The experimental data of Ref.~\cite{PhysRevLett.74.4627} are in good overall agreement with the three theoretical results, 
but indicate that the BGM ionization cross sections are probably too high. 
From this we can conclude that the present BGM calculations have the total ionization
probability under control at the level of $10 \%$ at low energies and better at energies above $100 \rm \ keV$.
The discrete excitation cross sections include populations of the $n=2,3,4$ shells, and for $n=2,3$ were checked for agreement witht the  QM-CCC 
results (not shown).

For the energy loss cross sections (shown in the right panel of Fig.~\ref{fig:Fig4}) 
BGM yields lower results than the SC-CCC approach~\cite{PhysRevA.92.022707} and the spline-basis TDSE solution of Ref.~\cite{PhysRevA.79.042901}
at the level of $5-10 \% $ and displays perhaps a slightly different shape.
This appears to be puzzling, given that we do not observe a shortfall in ionization, but rather the opposite. The only logical conclusion we can offer is that the average
energy in the BGM results is lower than the result from the close-coupling approach supplemented by a discrete wavepacket basis to provide an improved coverage
of the electronic continuum.
This seems to imply that
the average electron energy in the continuum is lower in the present work than in the TDSE solutions of 
Refs.~\cite{PhysRevA.79.042901,PhysRevA.92.022707}.

\subsubsection{Helium}
\label{ssec:he}

The helium target has received considerable attention from both the experimental and theoretical communities.
For the two-electron system it is possible to go beyond the independent-electron model advocated in the present work.
One approach to capture correlation effects, i.e., to take into account the large difference between the first and second ionization energies
of helium (24.59 and 54.42 eV respectively) is the independent event model, where separate calculations are performed for the `outer' and `inner'
$1s$ electrons of helium, and to combine the results.

Therefore, it is of interest to find out how the independent-particle model (IPM) fares in this regard. Its main problem is to extract a reliable double-ionization probability,
since the single-electron (IPM) ionization probability gives a decent estimate for one-fold ionization, but overestimates double ionization
when binomial statistics are applied. Another point of view is that the ionized single-particle density yields an accurate net probability, but the
separation into single- and double-ionization contributions is problematic.
Our results for antiproton-He collisions are therefore provided more for completeness, than as a serious competitor to true two-electron calculations.

\begin{figure}
\begin{center}$
\begin{array}{cc}
\resizebox{0.5\textwidth}{!}{\includegraphics{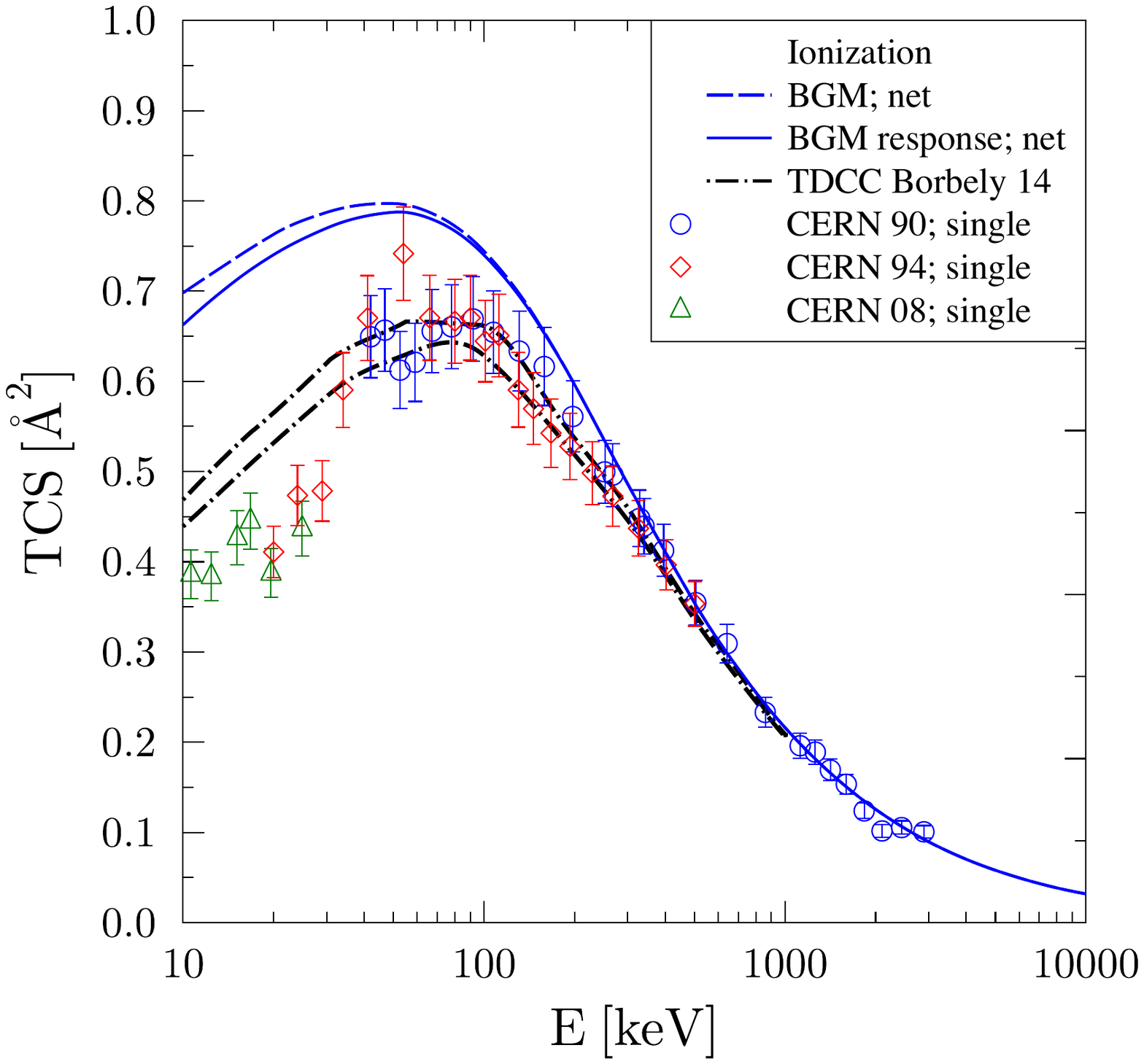}}{\hspace{-1.5 truecm}}&
\resizebox{0.5\textwidth}{!}{\includegraphics{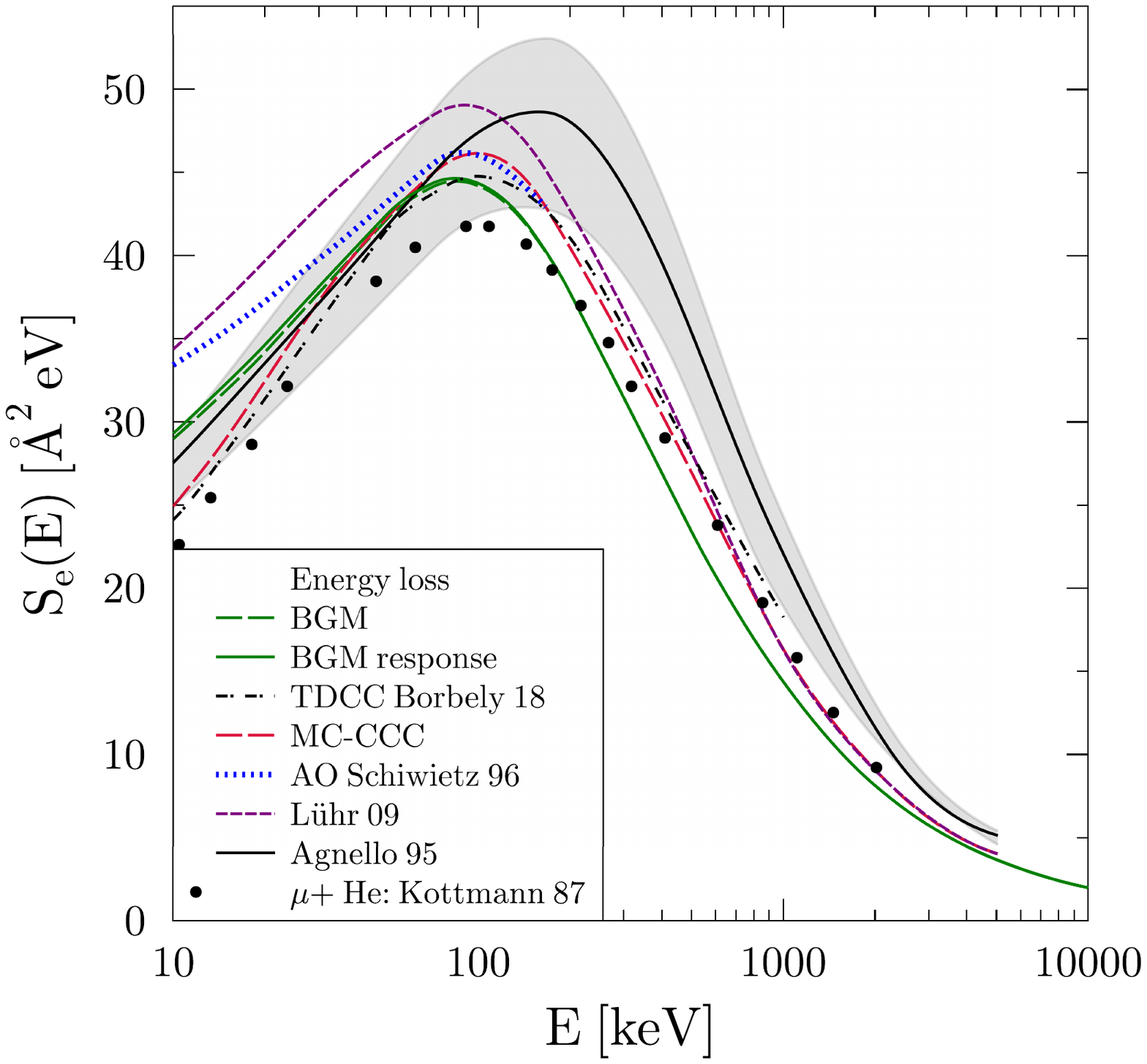}}
\end{array}$
{\vspace{-1.5 truecm}}
\caption{Energy dependence of ionization cross sections (left panel), and the energy loss cross sections (right panel)
for antiproton collisions with helium. The BGM results for net ionization are shown as a blue solid line (model with dynamical response)
and as a dashed blue line (static potential). The black chain curves are from Ref.~\cite{PhysRevA.90.052706}: the lower curve shows single
ionization, the upper curve net ionization.
The experimental data with error bars are for single ionization and were obtained during different runs at CERN\cite{PhysRevA.41.6536,Hvelplund_1994,PhysRevLett.101.043201}. 
In the right panel the BGM energy loss cross sections (the green solid line is with dynamical response, the green dashed line is with static
potential) are compared with the experimental data from Ref.~\cite{LODIRIZZINI2004,PhysRevLett.74.371} (black line with grey-shaded uncertainty estimate), and with the
experimental results for $\mu^-$ scattering~\cite{Kottmann} translated to the antiproton-equivalent collision energies. 
Other theoretical results: the black chain curve is from Ref.~\cite{PhysRevA.98.012707},
the purple short-dashed line from Ref.~\cite{PhysRevA.79.042901}, the red dashed line is the MC-CCC result from Ref.~\cite{PhysRevA.92.022707},
and the blue dotted line is the result from Ref.~\cite{Schiwietz_1996}.
}
   \label{fig:Fig5}
\end{center}
\end{figure}

The ionization cross sections are shown in the left panel of Fig.~\ref{fig:Fig5}. The experimental data for single ionization were described reasonably well
by the correlated finite element calculations of Borb\'ely {\it et al.}~\cite{PhysRevA.90.052706}, which make use of a discrete variable representation
for coupled-channel equations representing the two-electron wavefunction.
The experimental data from different runs do not form an entirely consistent picture to determine location and height of the maximum in the cross section precisely.
In Fig.~1 of Ref.~\cite{PhysRevA.90.052706} comparison is made of the single ionization cross section with many theoretical and experimental results over a wider
collision energy range, and it is found that the overestimation of experiment between impact energies of 10 and 40 keV is probably an experimental problem.
Good agreement of this observable is found there both with the QM-CCC calculations~\cite{Abdurakhmanov_2011}, and with full TDDFT calculations~\cite{PhysRevA.87.062507}.
For double ionization (cf. Fig.~2 of Ref.~\cite{PhysRevA.90.052706}) reasonable agreement is found with the CCC theoretical, as well as the experimental data with
a maximum cross section on the order of 0.02 \AA${^2}$ at an impact energy of $30-40 \ \rm keV$. This results in a difference between net and single ionization of
less than $10\%$. 

For the energy loss (shown in the right panel of Fig.~\ref{fig:Fig5} we can make the following observations:
While comparing the experimental data for antiproton impact~\cite{LODIRIZZINI2004,PhysRevLett.74.371} 
to the older $\mu^-$ impact data~\cite{Kottmann}, (given without
uncertainties, cf. also comments made in Ref.~\cite{PhysRevA.79.042901} about potential normalization issues), 
one finds that they agree at intermediate and high energies at the $10-20 \%$ level.
At high energies the sophisticated two-electron calculations, as well as the single-electron calculations
agree well with the muon scattering results. The present BGM calculations, on the other hand fall
short by a few percent in this range, even though the ionization cross sections do coincide with the
other theories. This would imply that in this energy regime single ionization dominates,
and that the BGM ionization is probably missing out on higher-energy electron contributions.

At intermediate energies most calculations seem to agree with a maximum at around 100 keV impact
energy and a value of about 45 \AA${}^2$eV. Somewhat higher results were reported in the single-active
electron work of Ref.~\cite{PhysRevA.79.042901}, which includes both single and double ionization.
In Ref.~\cite{PhysRevA.92.022707} a detailed analysis of two-electron effects was carried out at the level
of multi-configuration HF by
comparing MC-CCC to frozen-core calculations. It was shown that an increase in stopping power
at low to intermediate collision energies was obtained from the MC treatment. The more recent correlated two-electron work
of Ref.~\cite{PhysRevA.98.012707}, which is shown in Fig.~\ref{fig:Fig5}, is very close to the MC-CCC data,
and thus confirms the inadequacy of the single-active electron approach.

The present BGM data for energy loss show little difference whether dynamic response is included (solid green line)
or not (dashed line). It seems remarkable that the independent-electron model through the evaluation of loss by
the method introduced in this work is quite close to the sophisticated calculations of Refs.~\cite{PhysRevA.92.022707,PhysRevA.98.012707}
in the 40-100 keV energy range. The density-dependent evaluation method of energy loss can be used as an alternative
to the methods employed in the two-electron work of Refs.~\cite{PhysRevA.92.022707,PhysRevA.98.012707}, which is of interest
given the concerns raised of how true two-electron processes such as double ionization and excitation-ionization are treated
as sequential processes~\cite{PhysRevA.98.012707}.

\subsubsection{Neon}
\label{ssec:ne}

The neon atom represents a strong test for the independent particle model approach combined with the BGM for orbital propagation.
The observation of multiple ionization of rare gases such as Ne, Ar, Kr, Xe by ion impact led to the discovery that atomic physics was
not all about single-electron transitions. A number of collision systems was investigated theoretically: initially the field was dominated by
classical-trajectory approaches, but eventually the quantum treatment of the time-dependent many-electron problem that arises when using a
classical approach for the nuclear motion resulted in an understanding at what level one can understand the dynamics of these collisions.

For antiproton-neon collisions multiple ionization was observed experimentally by Paludan {\it et al.}~\cite{Paludan_1997b} in the 30-1000 keV energy regime, and single,
double and triple ionization cross sections $\sigma_i$ were reported. A theoretical explanation of these data was provided in Ref.~\cite{PhysRevA.61.012705}
using the approach advocated in the present work: screening effects are taken into account via a mean-field effective potential, which can be 
treated either as a static (no-response) model, or one can use the calculated net ionization probability as a function of time during the collision to
adjust the effective potential so that it acquires ionic character with fractional charge (dynamic response). 
The energy dependence of the experimental data~\cite{Paludan_1997b} is reproduced reasonably well for the three observed channels $i=1,2,3$
in Fig.~7 of Ref.~\cite{PhysRevA.61.012705}. Dynamical screening was found to be more important to describe the energy dependence of the multiple ionization cross sections
than for the net ionization cross section.

Compared to Ref.~\cite{PhysRevA.61.012705} we overestimate the net ionization cross section at energies below $100 \ \rm keV$ by 
about $8\%$. The reason for this is that in Ref.~\cite{PhysRevA.61.012705} solutions were obtained from a dynamic BGM ($W_p$ hierarchy), whereas in the current work we 
are restricted to a stationary BGM basis ($W_t$ hierarchy) in order to be able to calculate the energy loss using the Ehrenfest method. We also note that
the present dynamic response results are based on the approach introduced in Ref.~\cite{PhysRevA.62.042704}, while a simpler model was used in Ref.~\cite{PhysRevA.61.012705}.

\begin{figure}
\begin{center}$
\begin{array}{cc}
\resizebox{0.5\textwidth}{!}{\includegraphics{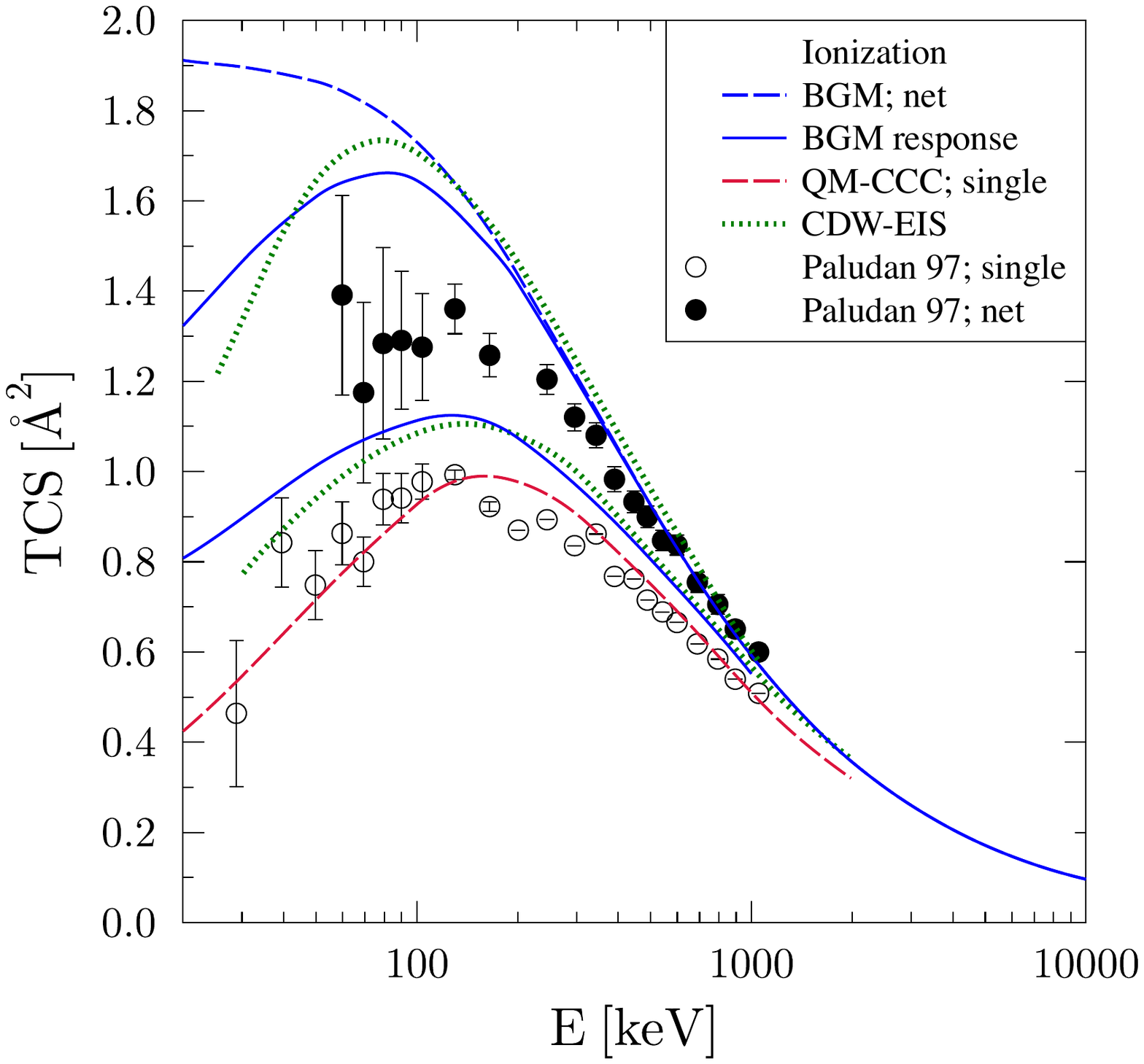}}{\hspace{-0.5 truecm}}&
\resizebox{0.5\textwidth}{!}{\includegraphics{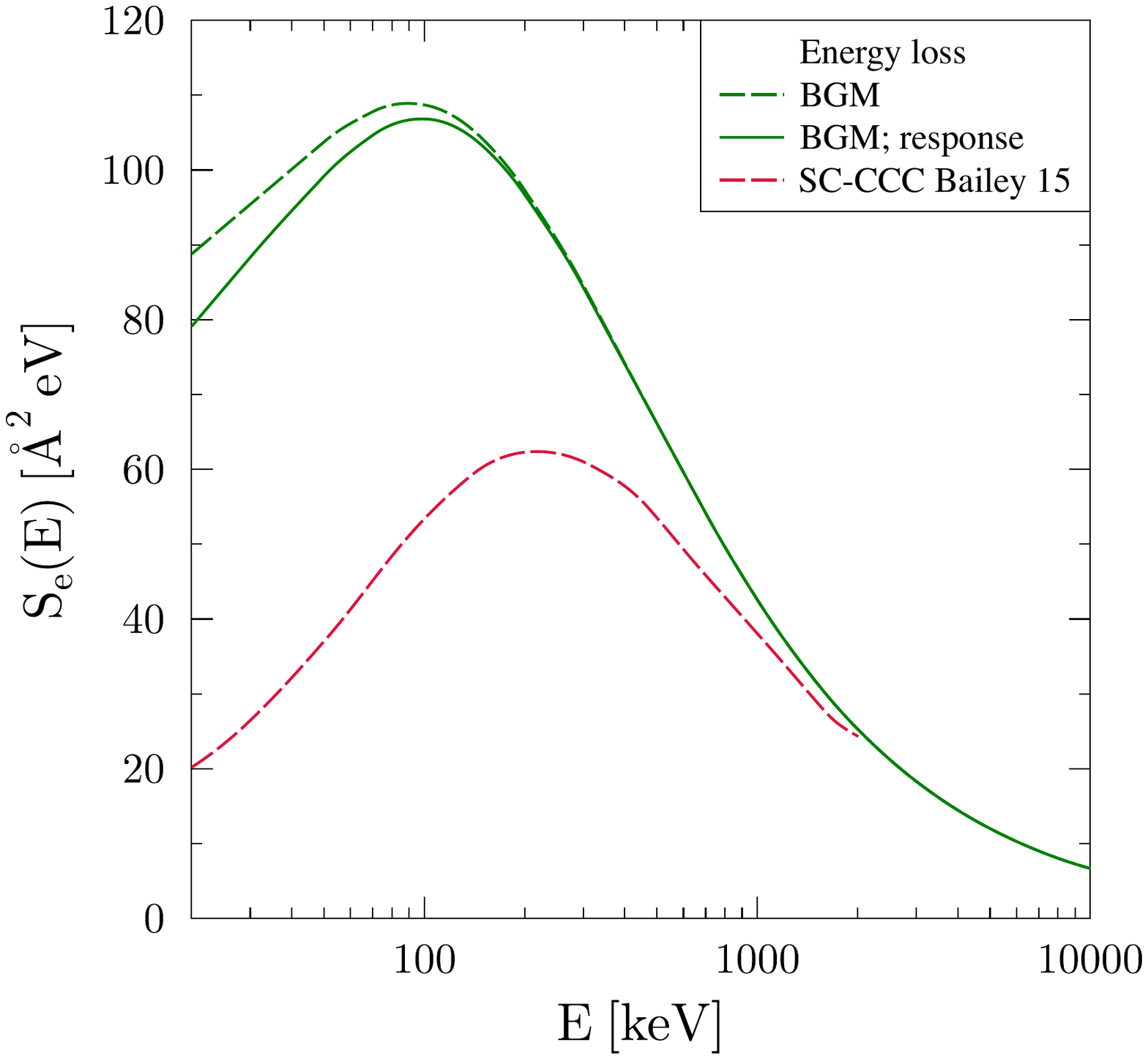}}
\end{array}$
{\vspace{-1.5 truecm}}
\caption{Energy dependence of ionization cross sections (left panel), and energy loss cross sections (right panel)
for antiproton collisions with neon.
The present $W_t$ hierarchy BGM results for ionization are shown as blue dashed and solid lines for static and dynamic response models respectively,
and the lower blue solid line shows the single ionization result from a binomial analysis. 
The  experimental net ionization cross sections are shown by solid circles with error bars and are derived from the single, double, and triple ionization data given in Ref.~\cite{Paludan_1997b} using eq.~(\ref{eq:netion}), while the open circles are the single-ionization cross sections $\sigma_1$ from Ref.~\cite{Paludan_1997b}.
The green dotted lines show the CDW-EIS results from Ref.~\cite{Montanari_2012}
(upper vs lower is net vs single ionization), while the red dashed line is the QM-CCC single-ionization cross section from Ref.~\cite{Abdurakhmanov_2011}.
Right panel: the green dashed and solid lines represent the energy loss calculated with static and with
dynamic response.
The red-dashed energy loss curve is from the SC-CCC calculation of Ref.~\cite{PhysRevA.92.022707}.}
   \label{fig:Fig6}
\end{center}
\end{figure}

While looking at the experimental data for ionization shown in the left panel of Fig.~\ref{fig:Fig6} we can clearly see the importance of multiple ionization
contributions.
The experimental net ionization cross section shown in Fig.~\ref{fig:Fig6} is based on the observed ionization cross sections $\sigma_i$ for
$i=1,2,3$ and calculated as
\begin{eqnarray}
\sigma_{\rm net} = \sum_{i=1}^3{ i \sigma_i} \ .
\label{eq:netion}
\end{eqnarray}
In principle, this is a low estimate, but we may assume that $i=4$ and higher ionization contributions were too small to be observed. The theoretical cross section
for  $\sigma_4$ was discussed in the context of {\it Continuum Distorted Wave with Eikonal Initial State} (CDW-EIS) calculations in Ref.~\cite{Montanari_2012}, 
and compared to electron impact ionization data.

The experimental single ionization cross section (open circles) is below the  CDW-EIS results from Ref.~\cite{Montanari_2012}, and also below the BGM results of Ref.~\cite{PhysRevA.61.012705}, as well as the present BGM calculations.
The QM-CCC calculation of Ref.~\cite{Abdurakhmanov_2011} ignores multi-electron contributions and is based on the propagation of the $2p$ shell only.
It apparently agrees best with the experimental values of $\sigma_1(E)$. 

When considering projectile energy loss it seems unreasonable to focus only on single-electron events, especially given the large discrepancy between the experimental single
and net ionization cross sections shown in the left panel of Fig.~\ref{fig:Fig6}. Our results for energy loss (right panel of Fig.~\ref{fig:Fig6})
demonstrate this clearly: the DFT-BGM calculations with or without dynamic response agree on the energy loss for energies above $100 \ \rm keV$; they may
overestimate the net ionization cross section at energies below $200 \ \rm keV$, if the data of Paludan {\it et al.}~\cite{Paludan_1997b} are taken at face value, but for higher
energies they do agree. Thus we find a large discrepancy with the QM-CCC results~\cite{Abdurakhmanov_2011} which we attribute to the contributions from 
the $2s$ shell of Ne, and many-electron processes in general, such as ionization-excitation.
Therefore, we conclude that for many-electron targets much has to be learned about energy loss, and measurements for neon (and heavier rare gases) would be most
welcome to settle the issue.

\subsubsection{Carbon, nitrogen, and oxygen}
\label{ssec:cno}
\begin{figure}
\begin{center}$
\begin{array}{ccc}
\resizebox{0.4\textwidth}{!}{\includegraphics{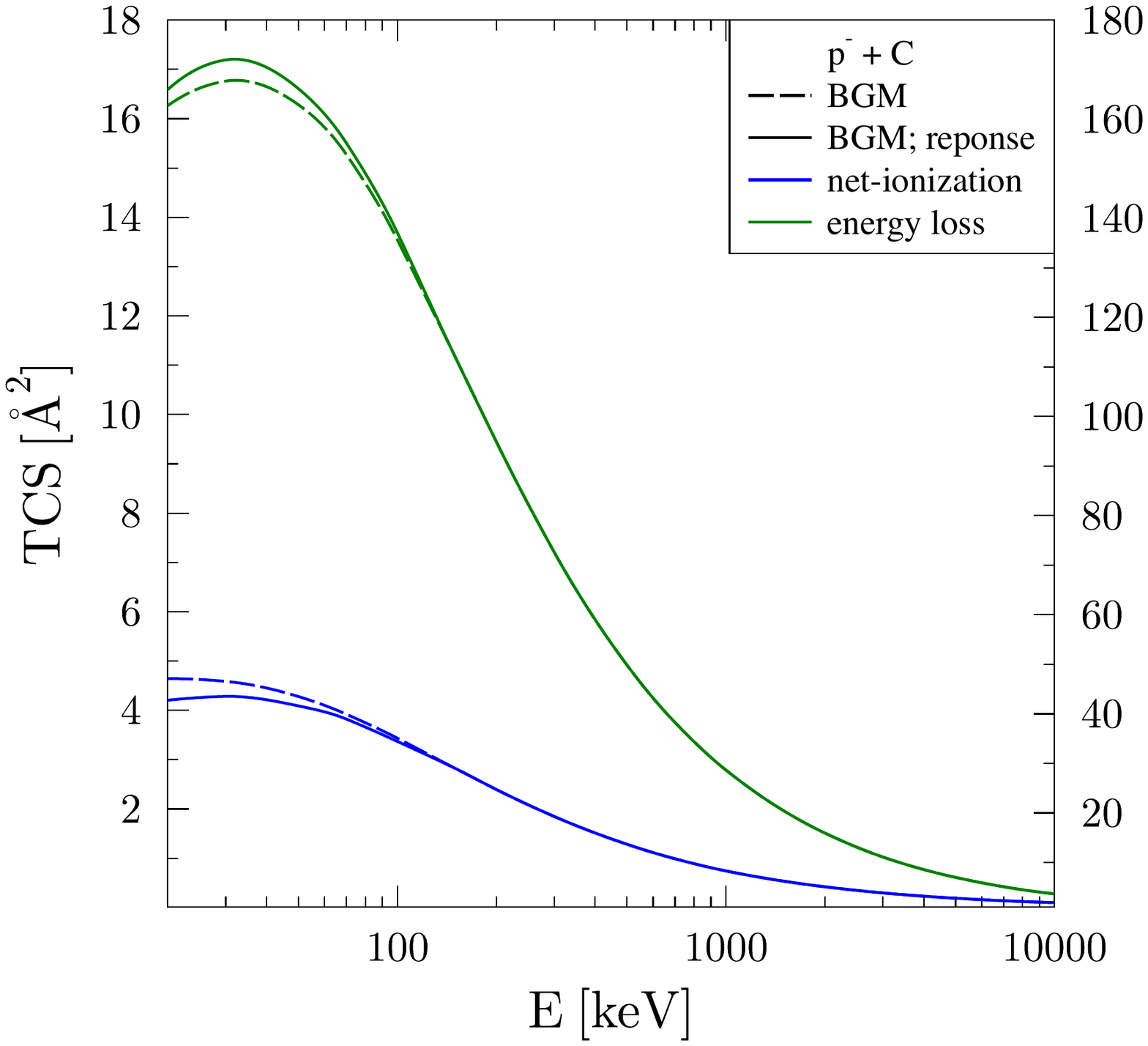}}{\hspace{-1.5 truecm}}&
\resizebox{0.4\textwidth}{!}{\includegraphics{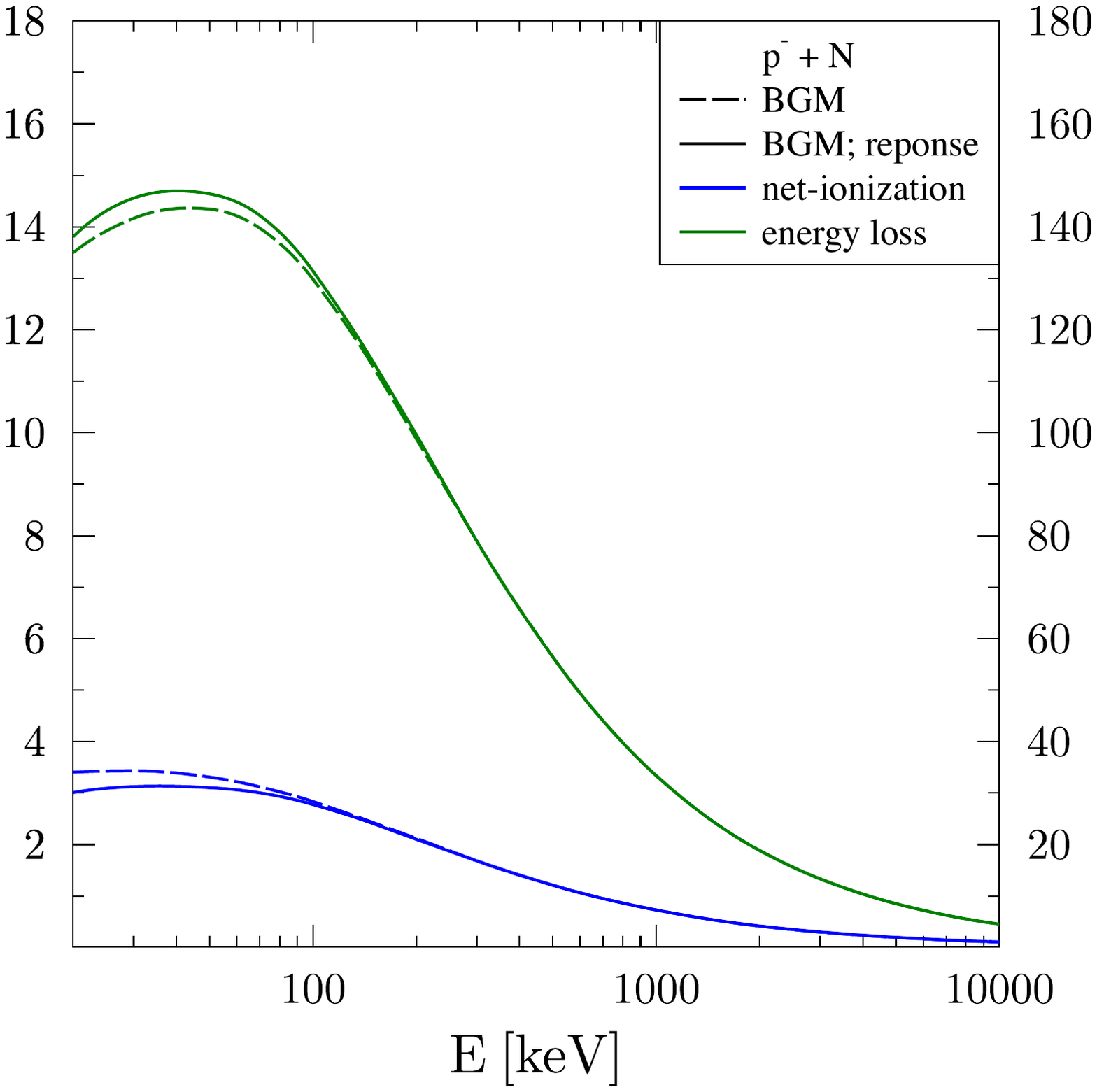}}{\hspace{-1.5 truecm}}&
\resizebox{0.4\textwidth}{!}{\includegraphics{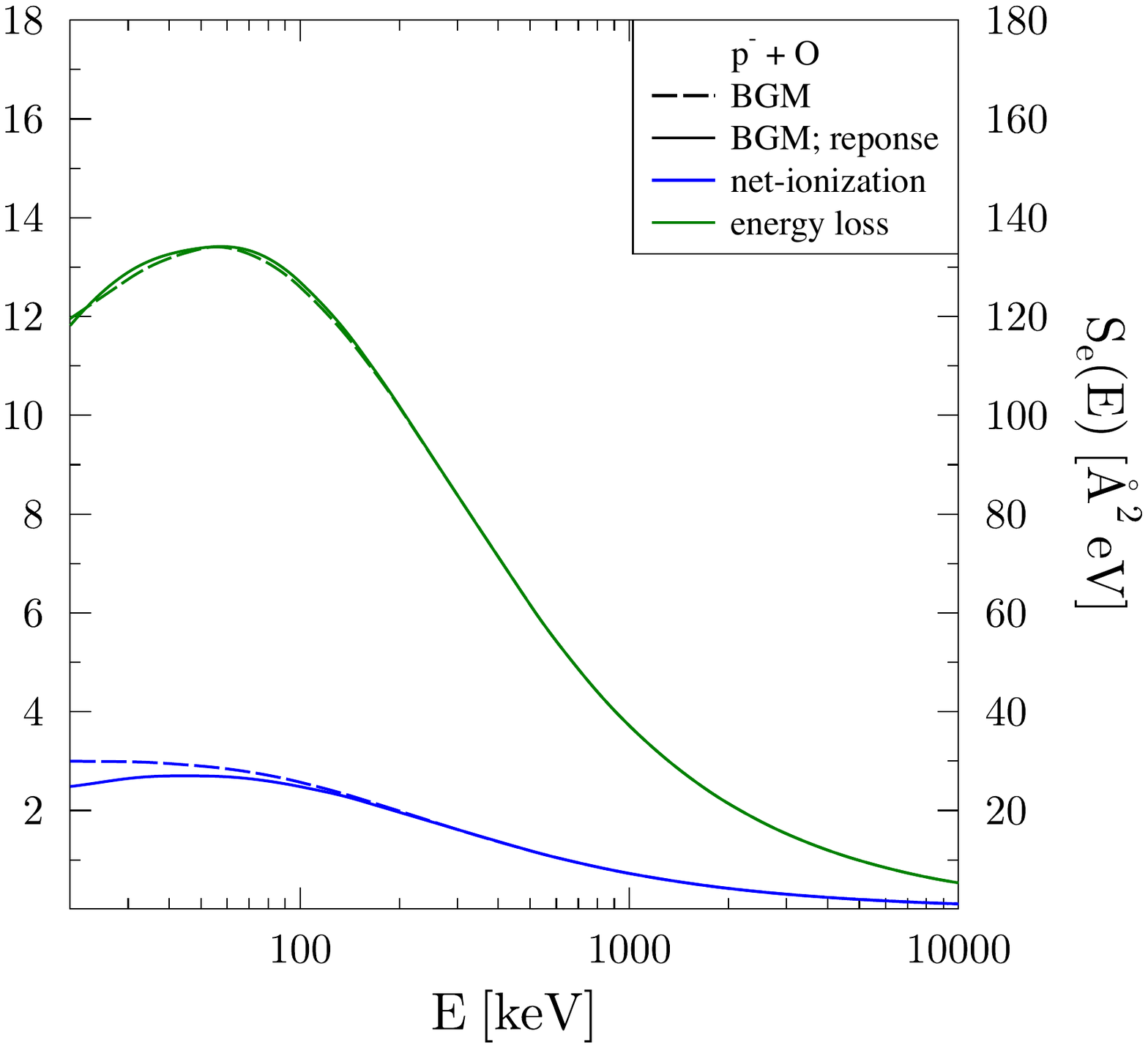}}
\end{array}$
{\vspace{-1.5 truecm}}
\caption{Energy dependence of cross sections for net ionization (shown in blue, lower curve pairs) and energy loss (shown in green using the scale on the right, upper curve pairs) for carbon (left panel), nitrogen (middle panel), and oxygen atoms (right panel).
Solid lines show results from BGM calculations with response, dashed lines without dynamical response.}
\label{fig:Fig7}
\end{center}
\end{figure}

In Fig.~\ref{fig:Fig7} we present results for second-row atoms of interest for independent-atom model calculations with molecular targets. The three atoms represent building blocks
for biologically relevant molecules, and have been investigated before to determine the strength of ionization~\cite{PhysRevA.101.062709} and capture 
processes~\cite{atoms8030059} in collisions with positively charged ions.
In analogy to helium and neon 
their atomic structure is described at the level of the optimized potential method, which represents the optimal Hartree-Fock equivalent treatment of DFT with the constraint
of a local potential. For antiproton impact the main ionization contributions come from $2p$ and $2s$ orbitals. The atoms are treated by using a spherically symmetric potential,
and the $2p_m$ magnetic sublevels are populated to equal fractions in order to obtain a spherically symmetric density for each neutral atom. 
The ionization energies obtained by
the method (as inferred from the $2p$ orbital energies via Koopmans' theorem)
are $11.72 \ \rm eV$ for C, $15.33 \ \rm eV$ for N, and $16.70 \ \rm eV$ for O. Except for oxygen these correspond closely to the experimental values of
$(11.26, 14.53, 13.62) \ \rm eV$ respectively.

In the dynamical response model the potential adjusts itself as a function of the time-dependent net ionization probability, which makes further ionization less probable (to a small degree for antiproton impact). The effect is clearly visible for all three targets at impact energies below $100 \ \rm keV$: the blue solid lines fall below the dashed ones, and the gap
widens as one lowers the impact energy.
The energy loss (green curves) shows a reversed pattern for carbon and nitrogen, but a negligible difference is found for oxygen between the two potential models.
Apparently, discrete excitations are increased by the inclusion of dynamic response, while ionization is reduced, and energy loss follows somewhat the 
trend observed for excitation cross sections (which are not shown).

The absolute values of net ionization decrease as one moves from carbon to oxygen targets, which is remarkable, since more electrons are available as one moves from
left to right in the figure. This is likely related to the fact that the $2p$ (and $2s$) orbital is bound more deeply as one moves through the sequence C, N, O, 
and therefore contributes less.

Comparing the energy loss results, which overall are very similar for both models (with and without dynamic response) we find that maxima occur between 30 and 80 keV
collision energy with a shift towards higher values within this range as one moves from carbon to oxygen.
While comparing the three atoms  
we note a slight decrease in the maximum heights of the cross sections, starting with 170 \ \AA${}^2 \rm  eV$ for carbon, 150 \ \AA${}^2 \rm eV$ for nitrogen,
and 135 \ \AA${}^2 \rm eV$ for oxygen.  This pattern is consistent with the behavior found for the net ionization
cross sections.

One can add to this sequence the results for neon (cf.~Fig.~\ref{fig:Fig6}), which in the BGM-OPM approach has an ionization energy of $23.15 \ \rm eV$, i.e., higher 
than the accepted value of $21.655 \ \rm eV$. The trend in the shift of the maximum to higher collision energies, and lowering of the maximum value continues
as observed in the sequence C, N, O, with
the maximum occurring at $100 \ \rm keV$, and a maximum value of less than 110 \ \AA${}^2 \rm eV$ in the case of neon, and a slightly reduced value for the
model with response.

\section{Conclusions}
\label{sec:conclusions}

\begin{figure}
\begin{center}$
\begin{array}{ccc}
\resizebox{0.4\textwidth}{!}{\includegraphics{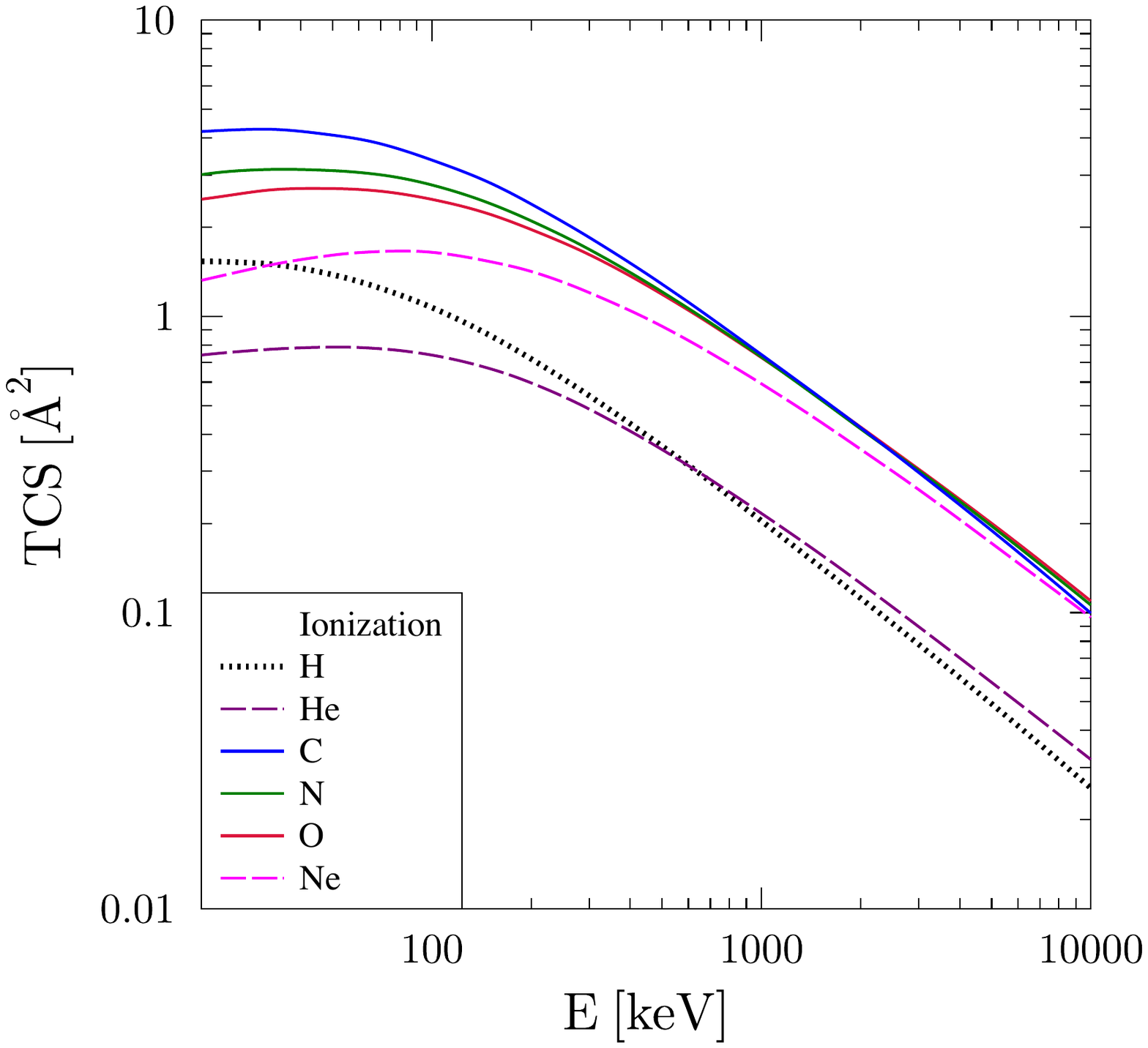}}{\hspace{-1.5 truecm}}&
\resizebox{0.4\textwidth}{!}{\includegraphics{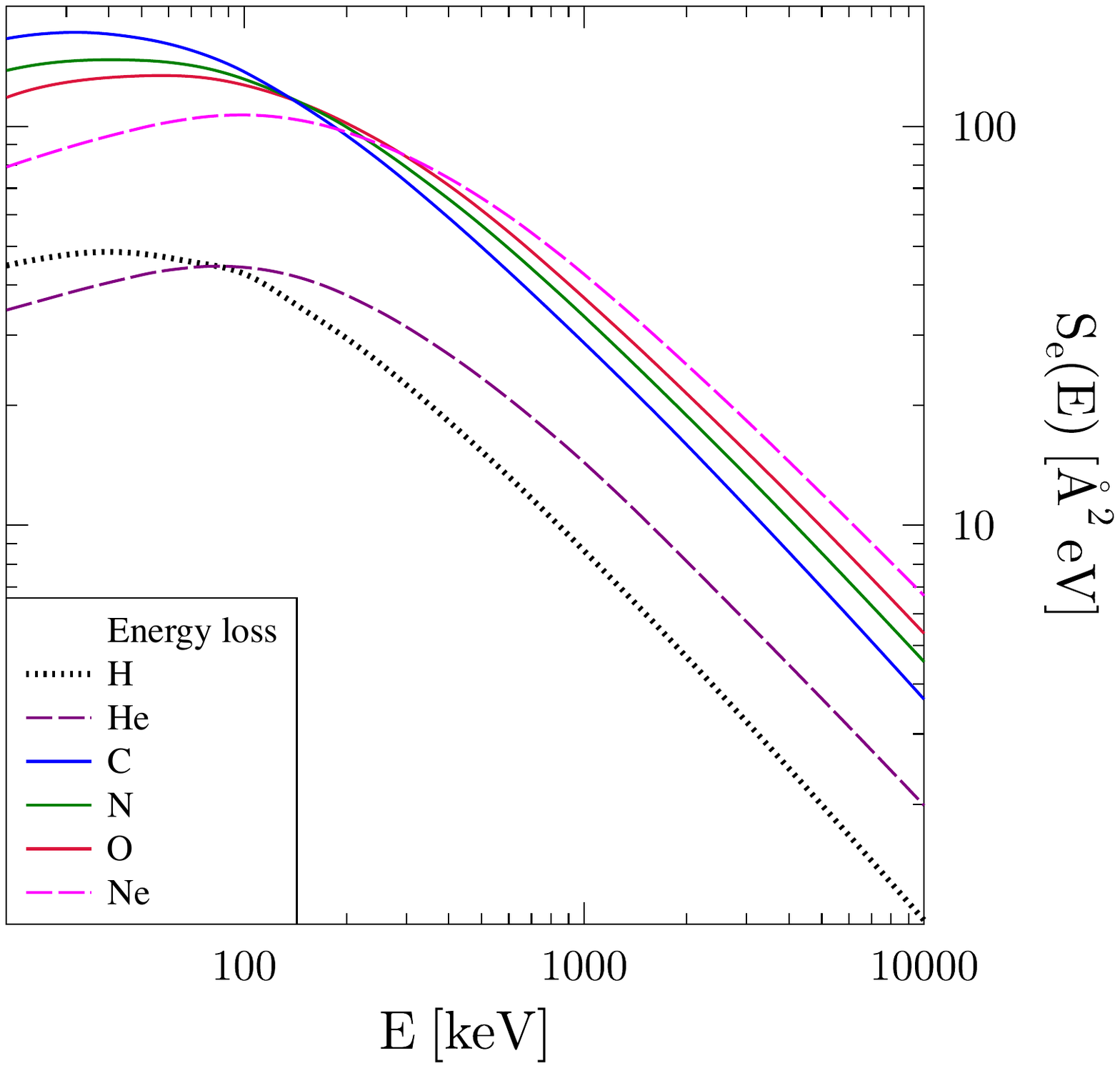}}{\hspace{-1.5 truecm}}&
\resizebox{0.4\textwidth}{!}{\includegraphics{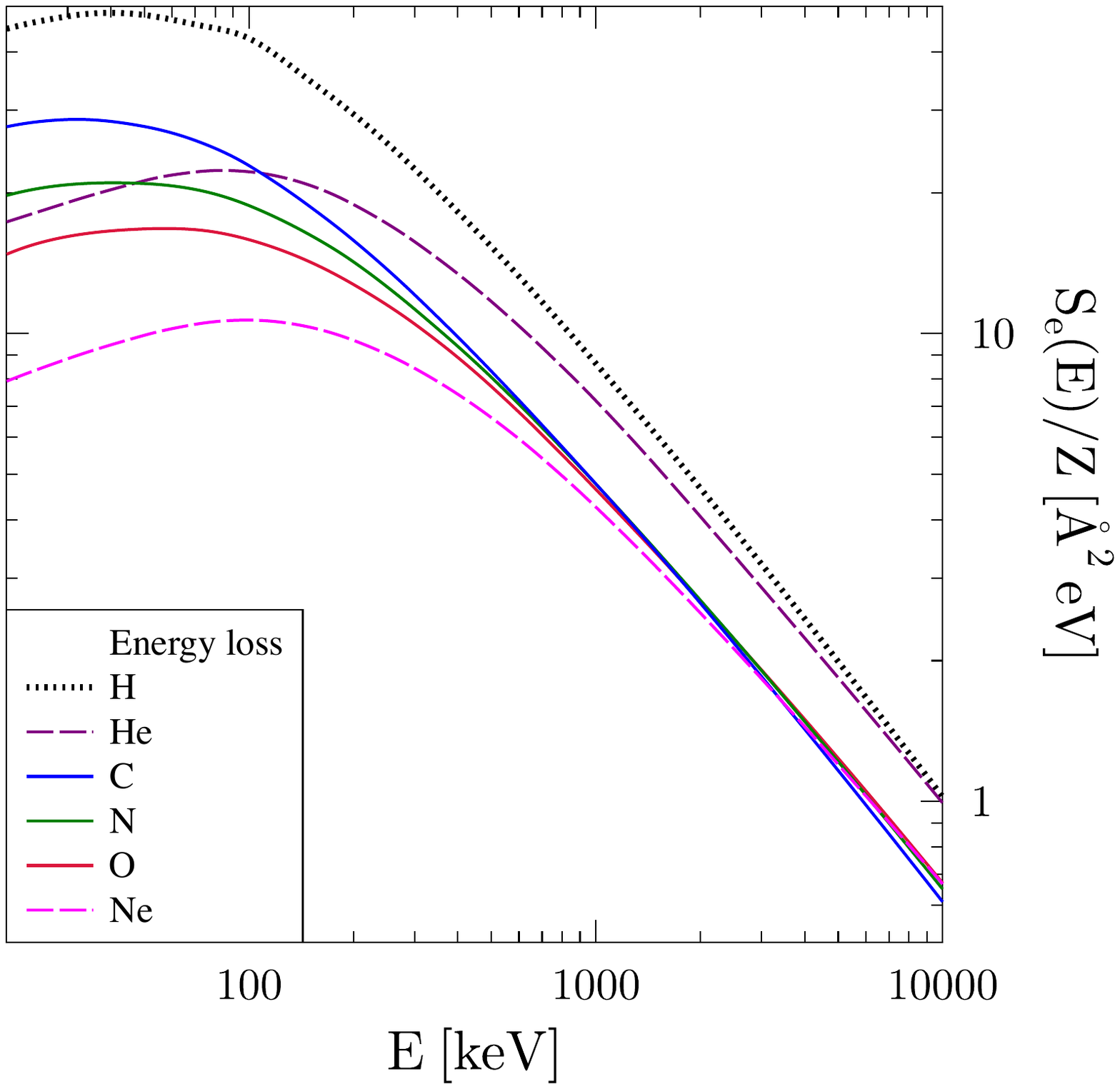}}
\end{array}$
{\vspace{-1.5 truecm}}
\caption{Comparison of cross sections for net ionization (left panel), and energy loss (middle panel) for 
the atoms H (dotted black), He (dashed purple, close to H at high energies), C (solid blue), N (solid green), O (solid red), and Ne (dashed magenta).
The right panel shows the results for the energy loss divided by the target electron number (or nuclear charge $Z$), see text.
The curves for C, N, O, Ne for energies $E<100 \ \rm keV$ are in this order from top to bottom.}
\label{fig:Fig8}
\end{center}
\end{figure}

In Fig.~\ref{fig:Fig8} we provide a summary of the presented BGM results within the dynamical response model for all atoms considered in the present work.
One can distinguish the intermediate energy regime (below 100 keV) from the higher energies. The net ionization cross sections below an impact energy of 100 keV
indicate that helium is hardest to ionize, followed by hydrogen and neon, while oxygen nitrogen and carbon have larger cross sections. This order is also
found for the energy loss cross section shown in the right panel with the exception of neon which rises higher towards the values for O, N, C.

At higher collision energies the ionization cross sections group themselves together into H and He with lower values, and the data for Ne, O, N, C merging into almost a
single curve. The energy loss, on the other hand, looks different at high energies. A well-defined sequence emerges, in which Ne rises to the top, and the trend in the sequence is with the number of electrons in the target, and with the (theoretical) ionization energies in the atomic sequence. The high-energy results are consistent with expectations based
on the Bethe-Born limit, cf., discussions in Refs.~\cite{hjl18,hjl19}.
An interesting question for comparison with experiment
would be to find out whether the large value of the theoretical ionization energy of oxygen is responsible for placing the O result between N and Ne.

In addition to the energy loss $S_e (E)$ defined in (\ref{eq:eloss}) one can also consider the mass stopping power
which is obtained by dividing $dE/dx$ by the mass density, i.e., $S_e(E)/M$, where
$M$ is the atomic mass. 
This macroscopic quantity is known for gas targets to become independent of pressure,
and neighboring atoms in the periodic table have comparable values.
Since we are showing results per atom in this work, following Bethe's work~\cite{Bethe1930} we 
prefer to display a related quantity, namely $S_e (E)/Z$, where $Z$ is the nuclear charge and equal to the number of target electrons.
This quantity is shown in the right panel of Fig.~\ref{fig:Fig8} and we find that the results for the atoms C, N, O, Ne fall 
on a common curve at high energies at about one third of the value for hydrogen.
At impact energies below 500 keV, however, the curves deviate and display a pattern similar to what
is seen in the left panel of Fig.~\ref{fig:Fig8} for the ionization cross sections.

The present work demonstrates that an evaluation of energy loss based on Ehrenfest's theorem which makes use of the density rather than complicated analysis of
wavefunctions represents a powerful tool. Using this tool we have obtained reasonable results for the one- and two-electron targets hydrogen and helium under antiproton
impact. A previous prediction for neon targets carried out over single-electron processes only within the CCC approach was shown to potentially underestimate
projectile energy loss by almost a factor of two at collision energies below 100 keV. BGM calculations of net ionization provide
results for a series of small atoms which are found in the second row of the periodic table for which definite predictions of energy loss are made.
In future work we plan to implement calculations for small molecules at the level of independent atom models~\cite{PhysRevA.101.062709,atoms8030059}.

\begin{acknowledgments}
We would like to thank the Center for Scientific Computing, University of Frankfurt for making their
High Performance Computing facilities available.
Financial support from the Natural Sciences and Engineering Research Council of Canada (NSERC) is gratefully acknowledged. 
\end{acknowledgments}

%

\bibliography{EnergyLoss}

\end{document}